%% file: main.tex
\documentclass[manuscript,screen, nonacm]{acmart}
\usepackage{multirow}
\usepackage{booktabs}
\usepackage[table]{xcolor}

\usepackage{makecell}

\usepackage{longtable}
\usepackage{array}
\usepackage{xurl} 
\DeclareUrlCommand\feat{\urlstyle{tt}}

\usepackage{array}
\newcolumntype{C}[1]{>{\centering\arraybackslash}p{#1}}

\usepackage{graphicx}

\newcommand{\blueicon}{%
  \raisebox{-0.5ex}{\includegraphics[height=3ex]{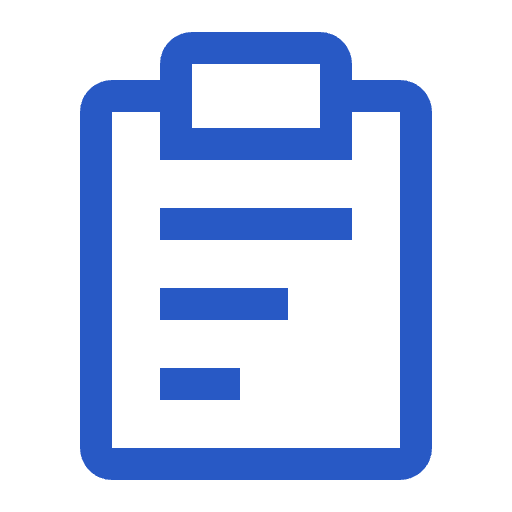}}%
}
\newcommand{\greenicon}{%
  \raisebox{-0.5ex}{\includegraphics[height=3ex]{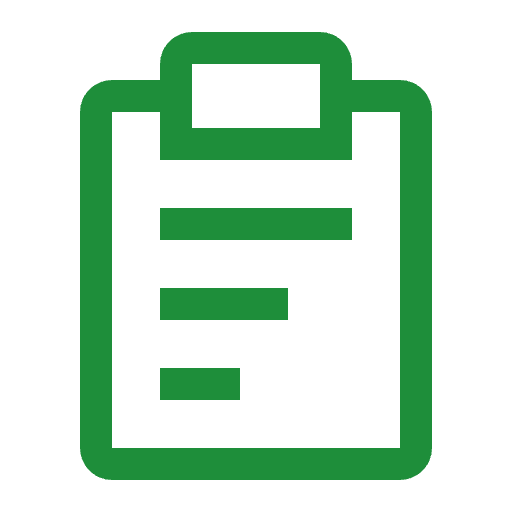}}%
}
\newcommand{\orangeicon}{%
  \raisebox{-0.5ex}{\includegraphics[height=3ex]{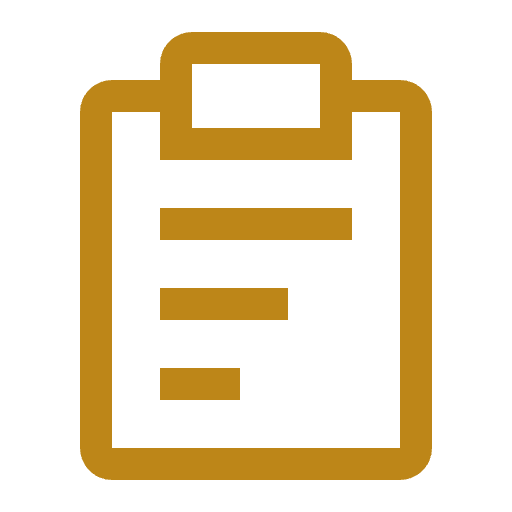}}%
}

\usepackage{color,soul}
\definecolor{lightblue}{RGB}{198, 226, 255} 
\sethlcolor{lightblue}

\usepackage[normalem]{ulem}

\newif{\ifhidecomments}
  \hidecommentstrue
\ifhidecomments
    \newcommand{\yishu}[1]{}

\else
    \newcommand{\yishu}[1]{\textbf{\small\sffamily{\textcolor{violet}{[#1 -- Yishu]}}}}
  \fi

\usepackage{listings}
\usepackage{xcolor}

\AtBeginDocument{%
  }

\setcopyright{none}
\begin{document}

% \title{Towards Zero-Training Mental Health Inference from Passive Sensing via Knowledge-Grounded Multi-Agent LLMs}

\title{OAK-Mind: Zero-Training Mental Health Change Inference from Passive Sensing via Knowledge-Grounded Multi-Agent LLMs}


\author{Yishu Ji}
\email{yishuji@cs.washington.edu}
\affiliation{%
  \department{Paul G. Allen School of Computer Science and Engineering}
  \institution{University of Washington}
  \city{Seattle}
  \country{United States}
}

\author{Zhihan Jiang}
% \email{zj2445@cumc.columbia.edu}
\affiliation{%
  \department{Biomedical Informatics}
  \institution{Columbia University}
  \city{New York}
  \state{New York}
  \country{United States}
}

\author{Will Ke Wang}
% \email{kw3215@cumc.columbia.edu}
\affiliation{%
  \department{Biomedical Informatics}
  \institution{Columbia University}
  \city{New York}
  \state{New York}
  \country{United States}
}

\author{Yuang Fan}
% \email{yf2676@columbia.edu}
\affiliation{%
  \institution{Columbia University}
  \city{New York}
  \state{New York}
  \country{United States}
}

\author{Anind K. Dey}
\affiliation{%
  \department{Information School}
  \institution{University of Washington}
  \city{Seattle}
  \state{Washington}
  \country{United States}
}

\author{Xuhai "Orson" Xu}
\affiliation{%
  \institution{Columbia University}
  \city{New York}
  \state{New York}
  \country{United States}
}

\renewcommand{\shortauthors}{Ji et al.}

\begin{abstract}

\input{sections/00_abstract}
\end{abstract}





\maketitle
\input{sections/01_introduction}

\input{sections/02_relatedwork}
\input{sections/03_methods}
\input{sections/04_evaluation}
\input{sections/05_results}
\input{sections/06_discussion}
\input{sections/07_conclusion}

\bibliographystyle{ACM-Reference-Format}
\bibliography{references}

% %
% %% If your work has an appendix, this is the place to put it.
\appendix

\include{sections/appx_kg}
\include{sections/appx_prompt}
\include{sections/appx_instance}
\include{sections/appx_features}

\end{document}
\endinput
%%
%% End of file `acmmanuscript.tex'.

%% file: sections/00_abstract.tex
Wearable and smartphone sensors continuously capture behavioral signals associated with mental health, yet inferring mental health outcomes from these signals has traditionally relied on supervised models trained on labeled data. Such methods require sufficient labeled examples to learn associations between behavior and mental health, while often providing limited insight into the evidence underlying their predictions. Large language models (LLMs) offer new opportunities to reason over behavioral evidence, but existing LLM-based approaches to passive sensing have largely continued to rely on task-specific fine-tuning or sensor-specific pretraining.
In this paper, we introduce \textbf{OAK-Mind}, a zer\textbf{\underline{O}}-training, multi-\textbf{\underline{A}}gent, \textbf{\underline{K}}nowledge-grounded LLM pipeline for inferring mental health changes from passive sensing data. 
The pipeline organizes sensing data into behavioral dimensions including sleep, physical activity, location mobility, and smartphone use, and orchestrates specialized agents grounded in mental health literature to interpret longitudinal behavioral changes. It synthesizes their assessments and applies questionnaire-based symptom rating to infer changes relative to an individual's baseline assessment with human-readable rationale grounded in observed behavioral patterns and domain knowledge. 
Across datasets and prediction tasks, while without task-specific training, OAK-Mind achieves comparable or better performance than the most predictive supervised baselines trained on additional labeled samples. It also demonstrate consistent improvement upon direct prompting across four tested LLM backbones. Ablation analyses support the benefits of combining behavioral interpretation with symptom rating, distributing analysis across domain-specific agents, and incorporating domain knowledge. Our findings highlight the potential of knowledge-grounded, multi-agent LLMs for training-free inference of mental health changes from passive sensing data.

%% file: sections/01_introduction.tex
\section{Introduction}

Smartphones and wearable devices continuously capture behavioral and physiological signals in everyday life, opening up opportunities to understand mental health beyond periodic self-report assessments and clinical visits~\cite{mohr2017personal}. Prior work in ubiquitous computing and mobile health has shown that patterns like sleep, physical activity, social interaction, and smartphone use are associated with mental health outcomes such as depression, anxiety, and stress \cite{amin_use_2025,canzian_trajectories_2015,wang_studentlife_2014}. 

Building on these associations, research has largely formulated mental health inference as a supervised learning problem by training computational models to map sensing-derived behavioral features with self-reported mental health outcomes \cite{saeb_mobile_2015, xu2022GLOBEMCrossDatasetGeneralization}. 
However, this paradigm depends on sufficiently large and high-quality passive sensing datasets with labeled mental health outcomes, which remain difficult to obtain \cite{xu2022GLOBEMCrossDatasetGeneralization, shen_passive_2025}. Moreover, those models provide limited insight into how observed behavioral changes support a particular mental health inference.

Recent advances in large language models (LLMs) offer a different opportunity for interpreting passive sensing data. Beyond prediction, LLMs can integrate contextual information and natural-language reasoning, thereby creating the potential to interpret behavioral patterns while providing human-readable rationale \cite{kimHealthLLMLargeLanguage2024c, cosentinoPersonalHealthLarge2024b}. Early studies have shown that LLM-based health inference can largely benefit from incorporating user information, temporal context, and health knowledge \cite{kimHealthLLMLargeLanguage2024c}. Yet general-purpose LLMs are not naturally structured to reason over sensing data, which are predominantly time-series, numerical, multi-modal, and provide limited semantic context for language-based reasoning \cite{zhang2025sensorlma}. Accordingly, state-of-the-art LLM-based approaches often relied on additional model adaptation, such as task-specific fine-tuning or sensor-specific pretraining \cite{cosentinoPersonalHealthLarge2024b, zhang2025sensorlma, xu2024mentalllm}. 
However, developing such specialized models requires suitable and sufficient training data and computational resources, as well as time and expertise for data preparation, training, and optimization.
These requirements can limit the practicality of model adaptation in sensing studies with constrained data or resources.

Motivated by these constraints, we develop \textbf{OAK-Mind}, a zer\textbf{\underline{O}}-training, multi-\textbf{\underline{A}}gent, \textbf{\underline{K}}nowledge-grounded  LLM system for inferring changes in an individual's mental health symptom severity (improves, remains stable, or worsens) over a longitudinal sensing period. The system structures inference into three stages via three agentic layers: First, a layer containing a set of behavioral agents analyzes longitudinal changes in sleep, physical activity, mobility, and smartphone use separately, with each agent guided by domain knowledge about how changes in its behavioral dimension may relate to mental health. Second, a cross-domain synthesis agent organizes evidence across these behavioral dimensions to form an integrated assessment of changes in symptom-related patterns. Finally, a symptom rating agent translates this assessment into a predicted change in the target mental health measure by grounding the inference in the corresponding questionnaire, while providing interpretable behavioral evidence to support the prediction. 

Our contributions are threefold:
\begin{itemize}
    \item \textbf{Zero-training mental health inference framework from passive sensing:} We introduce a training-free framework for inferring longitudinal changes in an individual's mental health without task-specific model training, fine-tuning, or labeled examples from the target dataset.
    \item \textbf{Domain knowledge harnessing and domain-specific analysis for grounded inference:} We design a multi-agent LLM harness that explicitly incorporates established knowledge about behavioral patterns and mental health as a reasoning scaffold, which grounds the inference in observable behavioral evidence.
    \item \textbf{A comprehensive validation:} We evaluate our approach across multiple datasets, mental health outcomes, and backbone LLMs. The results show that our approach achieved performance comparable to or better than state-of-the-art supervised machine-learning baselines without training on additional labels. The ablation studies validate that combining behavioral analysis and synthesis with questionnaire-based symptom rating outperformed either component alone, with further benefits from domain-specific agents design and knowledge harnessing.
\end{itemize} 

We envision that this work can highlight the potential of LLMs as knowledge-driven reasoning tools for passive sensing interpretation, while inspiring future research on data-efficient and interpretable approaches to understanding longitudinal human behavior and mental health change.

%% file: sections/02_relatedwork.tex
% Related Work
\section{Related Work}

\subsection{Passive Sensing and Mental Health Inference}
Behavioral health research has linked everyday behavior to mental health outcomes. Prospective evidence associates higher physical activity with lower depression risk~\cite{pearce2022physicalactivity}. Studies using wearable devices have also linked greater sleep regularity to lower subsequent depression and anxiety risk, and greater sleep variability and poorer sleep quality to worse depression outcomes~\cite{li2025sleepregularity,matcham2024sleeprelapse}.

Research in ubiquitous computing has examined these relationships using smartphones and wearable devices. Reviews have synthesized evidence on sensing-derived behavioral indicators of depressive symptoms~\cite{rohani2018correlations,deangel2022digital}. StudentLife and subsequent multimodal sensing studies investigated associations between sleep, physical activity, mobility, social behavior, and self-reported mental health~\cite{wang_studentlife_2014,Wang2018Tracking,moshe2021predicting}. Other studies examined associations of depressive symptoms with location variability, circadian movement, and phone use~\cite{saeb_mobile_2015,saeb2016location}, as well as prediction of depressive mood changes from mobility traces~\cite{canzian2015trajectories}.

Longitudinal studies have further examined how these relationships vary within individuals, across contexts, and over time. Stamatis et al. distinguished within-person and between-person associations with depression and anxiety, finding that time spent at home was associated with subsequent depressive symptoms~\cite{stamatis2024temporal}. Gyorda et al. found that phone-use associations with anxiety depended on where use occurred and varied across study years~\cite{Gyorda2024Longitudinal}. However, associations are not consistently observed: a prospective study of objectively measured smartphone and social-media use found generally small and nonsignificant within-person associations with subsequent depression, anxiety, and social isolation~\cite{smartphone2022distress}.

Longitudinal data resources pair passive sensing with repeated mental health assessments. In student populations, GLOBEM combines smartphone and wearable data across four annual 10-week studies, while the College Experience Study followed two cohorts over four years, including the COVID-19 pandemic~\cite{xu2022globemdataset,nepal2024capturing}. Beyond student populations, RADAR-MDD collected smartphone, wearable, and symptom data from people with recurrent major depressive disorder for up to two years~\cite{Matcham2022RADARMDD}.

\subsection{Machine Learning for Mental Health Inference from Passive Sensing}
Machine learning research has developed models to estimate depressive symptom severity and detect elevated symptoms from smartphone-derived behavioral features~\cite{saeb_mobile_2015,wahle2016mobile,opokuasare2021predicting}. Related work has used daily mobility features to predict social anxiety symptoms~\cite{boukhechba2018socialanxiety}. Other approaches incorporate symptom knowledge and behavioral context through behavioral proxies for depression-related symptoms and association-rule mining of contextually filtered features~\cite{Wang2018Tracking,xu2019routine}. Related work on loneliness also used association-rule mining to identify patterns involving location and time of day from smartphone and Fitbit data~\cite{Doryab2019loneliness}.

Beyond feature construction, researchers have examined individual differences and symptom changes over time. Personalized approaches include group-personalized regression that groups individuals with similar relationships between mobility features and depressive symptoms~\cite{palmius2018grouppersonalized}, and collaborative filtering that uses behavioral similarities among participants to identify personally relevant features across times of day and days of the week~\cite{xu2021Collaborative}. Addressing longitudinal outcomes, Chikersal et al.\ developed a feature extraction and selection approach using smartphone and fitness-tracker data to identify end-of-semester depressive symptoms and symptom worsening~\cite{chikersalDetectingDepressionPredicting2021b}.

Generalization remains a challenge for sensing-based mental health inference. Müller et al. found that GPS-based depression prediction performed substantially worse in a large, demographically heterogeneous sample than in a smaller student sample~\cite{muller2021depression}. GLOBEM benchmarked depression detection and domain generalization algorithms across longitudinal sensing datasets, finding substantial performance differences between within-dataset and cross-dataset evaluation settings~\cite{xu2022GLOBEMCrossDatasetGeneralization}.

\subsection{LLMs for Mental Health Inference}
Research has investigated LLM-based mental health inference from text, speech, and multimodal interview data. Mental-LLM evaluated zero-shot prompting, few-shot prompting, and instruction fine-tuning for mental health prediction from online text~\cite{xu2024mentalllm}, while MentaLLaMA used instruction tuning to generate mental health predictions and accompanying explanations from social-media posts~\cite{yang2024mentallama}. Beyond text, researchers have integrated acoustic landmarks with interview transcripts for depression detection~\cite{zhang2024acousticlandmarks}, and combined LLM-derived indicators from interview transcripts with facial-expression features to predict depressive symptom severity~\cite{sadeghi2024multimodal}.

For passive sensing data, studies have examined how LLMs can infer health outcomes from numerical behavioral and physiological measurements. Health-LLM evaluated zero-shot prompting, few-shot prompting, and instruction fine-tuning across wearable-based health prediction tasks, including mental health and sleep assessment~\cite{kimHealthLLMLargeLanguage2024c}. Focusing on longitudinal mental health inference, Yuan et al. evaluated changes in depression severity from digital phenotyping data collected from individuals experiencing major depressive episodes. They compared in-context learning with embedding-only and QLoRA fine-tuning and found that few-shot prompting and fine-tuning outperformed the traditional baselines evaluated in their study~\cite{yuan2026digitalphenotyping}. In related wellness tasks, HybridSense-LLM combined statistical features, wearable signal segments, and contextual self-reports in structured prompts to estimate stress, fatigue, readiness, and sleep quality, with prompting effectiveness varying across outcomes~\cite{Yu2026Hybridsense}.

Model adaptation has also been used to connect sensing data with language-based inference. PH-LLM fine-tuned Gemini for personalized sleep and fitness recommendations and trained a multimodal adapter to predict self-reported sleep outcomes from longitudinal wearable data~\cite{khasentino2025personala}. Other work has focused on sensor-language representations: SensorLLM aligned motion-sensor time series with textual trend descriptions for human activity recognition~\cite{li2025sensorllm}, while SensorLM used large-scale pretraining on wearable data paired with automatically generated descriptions, with evaluations covering zero-shot recognition, few-shot learning, and cross-modal retrieval~\cite{zhang2025sensorlma}.

Beyond outcome prediction, researchers have examined natural-language interpretations of sensing data. Englhardt et al.\ explored depression classification and interpretation of mobile behavioral trends, examining how practitioners combined generated analyses with their own expertise and knowledge of patients~\cite{Englhardt2024classtoclinial}. In a sleep-focused application, PhysioLLM combined wearable physiological data, contextual information, and statistical analyses to support users in understanding behavioral patterns and developing personal health goals~\cite{fangPhysioLLMSupportingPersonalized2024}.

\subsection{Agentic Interpretation of Personal Sensing Data}
Recent approaches have organized personal sensing interpretation into workflows that combine language-based reasoning with information retrieval. The Personal Health Insights Agent (PHIA) used iterative reasoning, code generation, and search to answer numerical and open-ended questions about wearable-derived behavioral health data~\cite{merrill2026phia}. SensorChat used question decomposition, sensor-data querying, and answer assembly to address qualitative and quantitative questions about long-term multimodal sensing data~\cite{yu2025sensorchat}.

Multi-agent approaches distribute analysis and interpretation across specialized roles. GLOSS assigned planning, information seeking, code generation, local and global sensemaking, and answer presentation to specialized agents, iteratively combining evidence across sensing streams to address open-ended questions~\cite{choube2025gloss}. Related work on personal health agents assigned data analysis, health-domain reasoning, and coaching to separate agents that jointly interpret wearable and health-record data and support personalized health interactions~\cite{heydari2025personalhealthagent}.

%% file: sections/03_methods.tex
\section{Methods}

\subsection{Problem Setting}

Mental health related symptoms can change over time, making longitudinal assessment important for understanding an individual's mental health. 
Assessing if symptoms have improved, remained stable, or worsened relative to an earlier evaluation helps characterize an individual's mental health trajectory. 
However, tracking these changes through repeated self-report assessments requires ongoing involvement from participants, which may be difficult to sustain.

We therefore study a setting in which an individual completes a mental health assessment at baseline ($T_1$) and contributes passive sensing data from their smartphone and wearable. Given the baseline assessment and sensing data from the periods preceding $T_1$ and a later time point $T_2$, the goal is to infer whether symptoms have improved, remained stable, or worsened, and to provide a rationale for the inference.

\subsubsection{Mental Health Measure}

We assess mental health in terms of depression and anxiety. The datasets used in our evaluation employ the PHQ-4~\cite{Kroenke2009PHQ4}, a four-item screening instrument for depression and anxiety. It consists of two subscales, the PHQ-2 for depression and the GAD-2 for anxiety, each with two items rated by how often the respondent experienced a symptom over the preceding two weeks (Table~\ref{tab:phq4}). We evaluate the two subscales separately.

\begin{table}[t]
\centering
\small
\renewcommand{\arraystretch}{1.2}
\caption{PHQ-4 items, subscales, and response options.
Item scores are summed within each subscale (range: 0--6).}
\label{tab:phq4}

\begin{tabular}{
  >{\raggedright\arraybackslash}p{2.5cm}
  >{\raggedright\arraybackslash}p{6.5cm}
  >{\raggedright\arraybackslash}p{3.6cm}}
\toprule
\textbf{Subscale} &
\textbf{Item} \newline \textit{"In past 2 weeks, how often have you been bothered by:"} &
\textbf{Response options} \\
\midrule

\multirow{2}{*}{Depression (PHQ-2)}
  & - Feeling down, depressed, or hopeless
  & \multirow{4}{=}{%
      \raggedright
      0 = Not at all\newline
      1 = Several days\newline
      2 = More than half the days\newline
      3 = Nearly every day
    } \\
  & - Little interest or pleasure in doing things
  & \\

\cmidrule(lr){1-2}

\multirow{2}{*}{Anxiety (GAD-2)}
  & - Feeling nervous, anxious, or on edge
  & \\
  & - Not being able to stop or control worrying
  & \\

\bottomrule
\end{tabular}
\end{table}

\subsubsection{Problem Formulation}
For individual $i$, let $s_{1,i}$ and $s_{2,i}$ denote the subscale scores at
$T_1$ ("Baseline") and $T_2$ ("Follow-up"), where $T_1$ is earlier than $T_2$, and let $X_{1,i}$ and $X_{2,i}$ denote the passive sensing
features aggregated over the windows preceding each time point. We form the comparison of scores between $T_1$ and $T_2$ as a 3-class classification problem; that is, with
$\Delta_i = s_{2,i} - s_{1,i}$, and a higher PHQ score indicates a more severe symptom, the change labels are:
\[
y_i =
\begin{cases}
\textsc{Improved} & \text{if } \Delta_i < 0, \\
\textsc{Stable}   & \text{if } \Delta_i = 0, \\
\textsc{Worsened} & \text{if } \Delta_i > 0.
\end{cases}
\]
OAK-Mind compares the two sensing windows in light of the baseline score and outputs a predicted label with a natural-language rationale:
\[
  f_i: \bigl(X_{1,i},\, X_{2,i},\, {s}_{1,i}\bigr) \;\mapsto\; (\hat{y}_i,\, r_i),
\]
where $\hat{y}_{i}$ is the predicted value of ${y}_{i}$, and $r_i$ is the rationale of this prediction. The baseline input $s_{1,i}$ includes both the subscale score and its item-level scores. 
OAK-Mind implements $f$ as the process of individual sensemaking, which routes through a sequence of knowledge-grounded agents that produce interpretable intermediate narratives to understand the patterns. The detailed pipeline is presented in Section~\ref{sec:system_design}.

\subsubsection{Zero-training setting}
OAK-Mind is built on off-the-shelf LLMs, where no model parameters are updated, and no labeled examples from other participants are used, either as training data or as in-context demonstrations. The only label-bearing input is the participant's own $T_1$ score. This contrasts with the common requirements in prior work: population-level supervised models, which learn from sensing and survey pairs across many individuals, or the personal models, which require measures for the same individual over an extended period.

% \subsubsection{Temporal gap}
% In this work, we require $T_2 - T_1 \ge 50$ days examine changes over a longer interval, given that the PHQ-4 assesses symptoms over the preceding two weeks.

\subsubsection{Temporal gap}
We require $T_2 - T_1 \ge 50$ days to examine symptom changes over an interval that is substantially longer than the two-week period of the PHQ-4. This threshold is chosen as it balances temporal separation between assessments while retaining a sufficient number of eligible assessment pairs for analysis in all the datasets we evaluate.

\subsection{System Design}
\label{sec:system_design}

\begin{figure}
    \centering
    \includegraphics[width=1\linewidth]{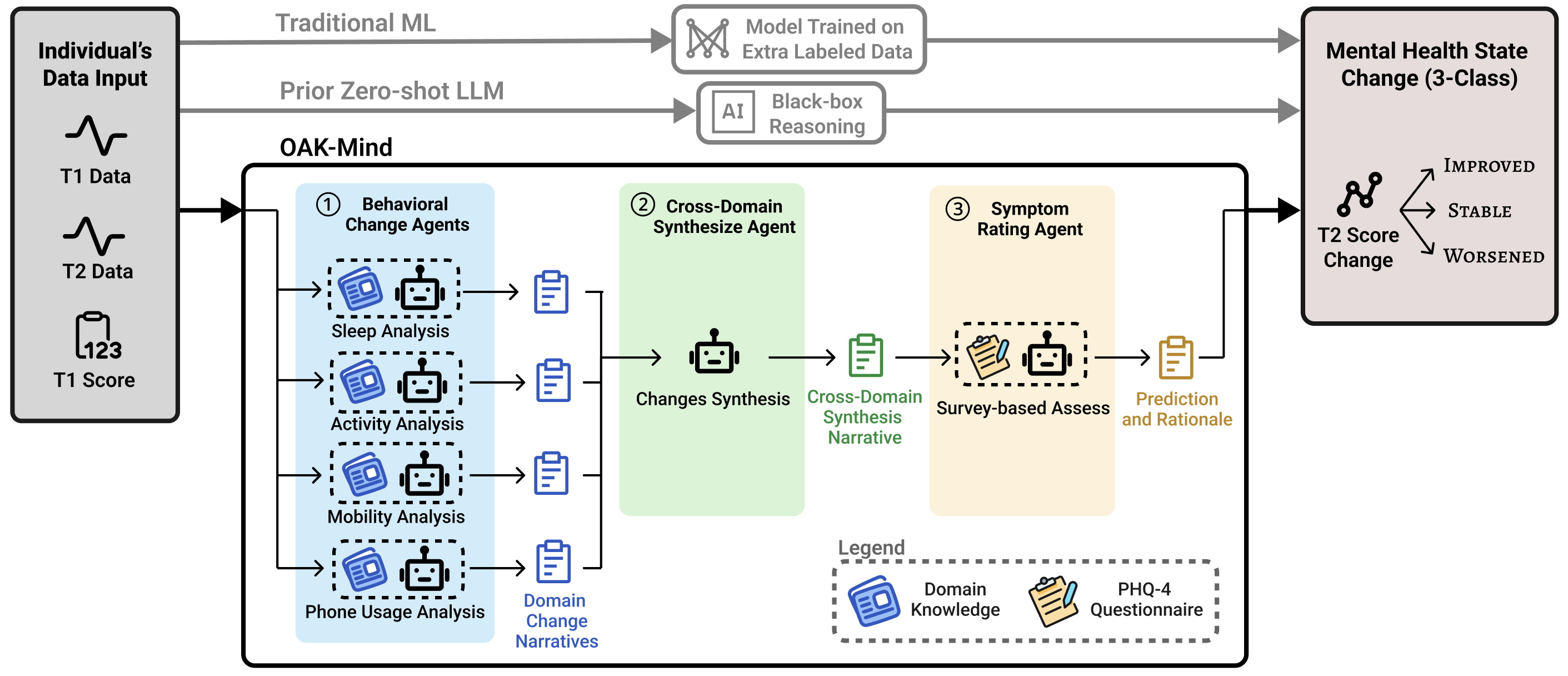}
    \caption{The overall design of the OAK-Mind pipeline}
    \label{fig:pipline}
\end{figure}

% As Figure ~\ref{fig:pipline} shows, OAK-Mind is composed of 3 layers of agentic design: The Behavioral Change Layer (\textcircled{1}), the Cross-Domain Synthesis Layer (\textcircled{2}), and the Symptom Rating Layer (\textcircled{3}).
As Figure ~\ref{fig:pipline} shows, OAK-Mind organizes inference from passive sensing with three layers of agents: Behavioral Change Agents (\textcircled{1}), a Cross-Domain Synthesis Agent (\textcircled{2}), and a Symptom Rating Agent (\textcircled{3}). 
All agents are instantiated from a single backbone LLM and specialized through prompting without any task-specific training. Agents communicate through natural-language narratives, so that every intermediate step and the final prediction comes with a human-readable rationale for potential tracking. 
Table ~\ref{tab:agents} demonstrates a summary on the specific roles, inputs, outputs of each agentic layers. A complete set of prompts used by OAK-Mind for one dataset–task configuration is provided in Appendix~\ref{app:prompt}.

\subsubsection{Behavioral Change Agents}
The first layer consists of a set of four agents, each of them is responsible for one
behavioral domain.
All four agents share a common prompt template, instantiated with their specific domain knowledge (Table~\ref{tab:domain_knowledge}). Each agent receives the participant’s baseline and follow-up sensing features for its assigned domain. Following prior work on zero-shot health prediction from passive sensing data~\cite{kimHealthLLMLargeLanguage2024c}, we convert feature names into short readable descriptions and represent each feature as a chronological sequence of daily values within the corresponding sensing window. 
For example, a snippet from one participant's input reads:

\begin{quote}
\small
\texttt{The daily record of a user's sensor readings in the
recent 14 days (oldest to newest) shows:
Time After Wake Up (min):
\{3.0, 0.0, 3.0, 0.0, 0.0, 0.0, 0.0, 3.0, 1.0, 0.0,
0.0, 0.0, 0.0, 1.0\}; \ldots}
\end{quote}

With the domain knowledge grounding, the agent compares these sequences to describe how behavioral patterns change from $T_1$ to $T_2$ and produces a narrative on the behavioral changes with a directional judgment.

How domains were selected and domain knowledge was extracted are described as follows:
\paragraph{Domain Selection.} 
Mobile and wearable devices capture a wide range of modalities, but not all are equally informative about mental health. For OAK-Mind, we selected four behavioral domains: \textbf{sleep}, \textbf{physical activity} (denoted as "\textbf{activity}"), \textbf{location mobility} (denoted as "\textbf{mobility}"), and \textbf{phone usage}. These domains are selected as they are associated with mental health and can be collected passively. Sleep and physical activity are reflected in diagnostic criteria for
depression~\cite{apa2013dsm5,Kroenke2001PHQ9,Wang2018Tracking}, while prior digital health research has linked all four domains to depressive symptoms~\cite{deangel2022digital,saeb_mobile_2015,canzian_trajectories_2015}.
They are also commonly captured in longitudinal sensing datasets~\cite{xu2022globemdataset,nepal2024capturing}.

\begin{table}[t]
\centering
\small
\caption{Domain knowledge embedded in each Behavioral Change Agent for the depression and anxiety tasks.}
\label{tab:domain_knowledge}
\begin{tabular}{p{1.8cm} p{1.6cm} p{9.4cm}}
\toprule
\textbf{Domain} & \textbf{Task} & \textbf{Domain Knowledge Summary} \\
\midrule
\multirow{2}{*}{Sleep}
  & \multirow{1}{*}{Depression} & Improvement in depression-related sleep often appears as easing of disruptions, reduced variability, fewer extreme nights, or more predictable timing. \\
\cmidrule(lr){2-3}
  & \multirow{1}{*}{Anxiety}    & Higher anxiety often appears alongside shorter sleep duration, and might coincide with reduced sleep efficiency, more time spent awake after sleep onset, poorer overall sleep quality and more disturbed sleep. \\
\midrule
\multirow{2}{*}{Activity}
  & \multirow{1}{*}{Depression} & Improvement in anhedonia-related activity patterns often appears as reduced inactivity, more consistent daily movement, or partial re-engagement rather than high activity levels. \\
\cmidrule(lr){2-3}
  & \multirow{1}{*}{Anxiety}    & More activity and higher step counts generally linked to lower anxiety. More sedentary time and lower fulfillment of activity guidelines are associated with higher anxiety. \\
\midrule
\multirow{2}{*}{Mobility}
  & \multirow{1}{*}{Depression} & Depressive symptom improvement may involve reduced withdrawal, slightly increased outings, or more consistent mobility patterns, even if overall movement remains limited. \\
\cmidrule(lr){2-3}
  & \multirow{1}{*}{Anxiety}    & Lower diversity of places visited (more repetitive, less varied locations) and more time spent at home may be associated with higher anxiety\\
\midrule
\multirow{2}{*}{Phone Usage}
  & \multirow{1}{*}{Depression} & Easing of depressive symptoms may appear as reduced disengagement, more regular phone use, or modest increases in interaction. \\
\cmidrule(lr){2-3}
  & \multirow{1}{*}{Anxiety}    & Higher screen time, especially during nighttime,  more frequent phone unlocking is associated with higher anxiety. More text messaging may indicate lower anxiety but not always. \\
\bottomrule
\end{tabular}
\end{table}

\paragraph{Domain Knowledge Extraction.}
To ground each agent in prior evidence, we retrieved relevant
relationships from a mental health knowledge graph constructed using LLMs to extract entities and relationships from research literature and clinical reference documents relevant to mental health and passive sensing (Appendix \ref{app:kg}).
For each behavioral domain and target outcome (depression or anxiety), we defined sets of related keywords (e.g., for sleep: \textit{sleep duration}, \textit{sleep efficiency}, \textit{bedtime}, \textit{wake}...). We retained relationships when their entity names matched a domain keyword and an outcome keyword, respectively, and their relation type indicated a direct association (e.g., \textit{predicts}, \textit{correlates with}, or \textit{associated with}). We then summarized the retained relationships as natural-language statements and included them in the corresponding agent's prompt.

\subsubsection{Cross-Domain Synthesis Agent}
Evidence from any single domain is noisy and can be ambiguous or demonstrate distinct mental state change tendency with another domains. The Cross-Domain Synthesis Agent therefore integrates the four
domain narratives, together with the baseline score $s_{1,i}$, into a holistic assessment of symptom change. It evaluates whether symptom-related signals have eased, persisted, or worsened, and weighs
whether the domains provide consistent or conflicting evidence. The baseline score lets the agent additionally account for the room for change; for instance, a participant near the scale floor has little room to improve. The agent outputs a natural language narrative concluding the symptoms the participant demonstrated and relevant reasoning on if they have likely improved, remained stable, or worsened.

\begin{table}[t]
\centering
\footnotesize
\renewcommand{\arraystretch}{1.15}
\caption{Role, prompt input, and output of each agent in OAK-Mind (shown for Depression Task). Quoted text shows the instructions given to each agent; \texttt{<>} denotes the information an agent receives or produces. A set of full prompts is provided in Appendix~\ref{app:prompt}.}
\label{tab:agents}
\begin{tabular}{
  >{\raggedright\arraybackslash}p{3.3cm}
  >{\raggedright\arraybackslash}p{7.2cm}
  >{\raggedright\arraybackslash}p{3.8cm}}
\toprule
\textbf{Agent \& Role} & \textbf{Input Prompt Structure} & \textbf{Output} \\
\midrule
\textbf{\textcircled{1} Behavioral Change Agents}: Compare a participant's baseline and follow-up behavioral data
  & \textbf{Role:} "You are a \texttt{<domain>} behavior expert analyzing wearable-derived \texttt{<domain>} features and their association with depression." \newline
    \textbf{Domain knowledge:} \texttt{<Domain knowledge summary>}, 
    \newline
    \textbf{Data:} \texttt{<Baseline features $X_{1,i}$>}; \texttt{<Follow-up features $X_{2,i}$>}. \newline
    \textbf{Task:} "Compare baseline vs.\ follow-up \texttt{<domain>} data and describe how symptom-related \texttt{<domain>} patterns evolve..." 
  &   \texttt{<\blueicon Domain change narrative>} $\times$ 4\\
\midrule
\textbf{\textcircled{2} Cross-Domain Synthesis Agent}: Integrate evidence across domains
  & \textbf{Role:} "You are synthesizing behavioral changes across multiple domains to assess depression state change." \newline
    \textbf{Data:}  \texttt{<\blueicon Domain change narrative>} $\times$ 4; \texttt{<baseline subscale score ${s}_{1,i}$>}. \newline
    \textbf{Task:} "Assess whether symptom-related signals have eased, persisted, or worsened..."
  & \texttt{<\greenicon Cross-domain synthesis narrative>} \\
\midrule
\textbf{\textcircled{3} Symptom Rating Agent}: Map the synthesis onto questionnaire and infer the direction of change
  & \textbf{Role:} "You are a mental health assessment expert that rates the individual's  mental state changes." \newline
  \textbf{Questionnaire: } \texttt{<PHQ-4 items and response scale>} \newline
  \textbf{Data:} \texttt{<\greenicon Cross-domain synthesis narrative>}; \texttt{<baseline subscale score ${s}_{1,i}$>} \newline
    \textbf{Task:} "Analyze each symptom dimension based on the behavioral changes. Then make an overall judgment..."
  & \texttt{<\orangeicon Predicted change $\hat{y}_i$ and Rationale $r_i$>} \\
\bottomrule
\end{tabular}
\end{table}

\subsubsection{Symptom Rating Agent}
The Symptom Rating Agent produces the final prediction by following the structure and scoring procedure of the relevant PHQ-4 subscale, grounding the inference in the questionnaire's symptom dimensions. Given the synthesis narrative and a participant's baseline item scores, it estimates a follow-up score for each of the two items based on the described behavioral changes. It then sums these estimates and compares the resulting subscale score with the baseline score to classify symptom change. The output includes the predicted label $\hat{y}_i$ (\textsc{Improved}, \textsc{Stable}, or \textsc{Worsened}) and a brief rationale $r_i$ addressing both symptom dimensions as a human-readable justification.

\begin{figure}
    \centering
    \includegraphics[width=0.98\linewidth]{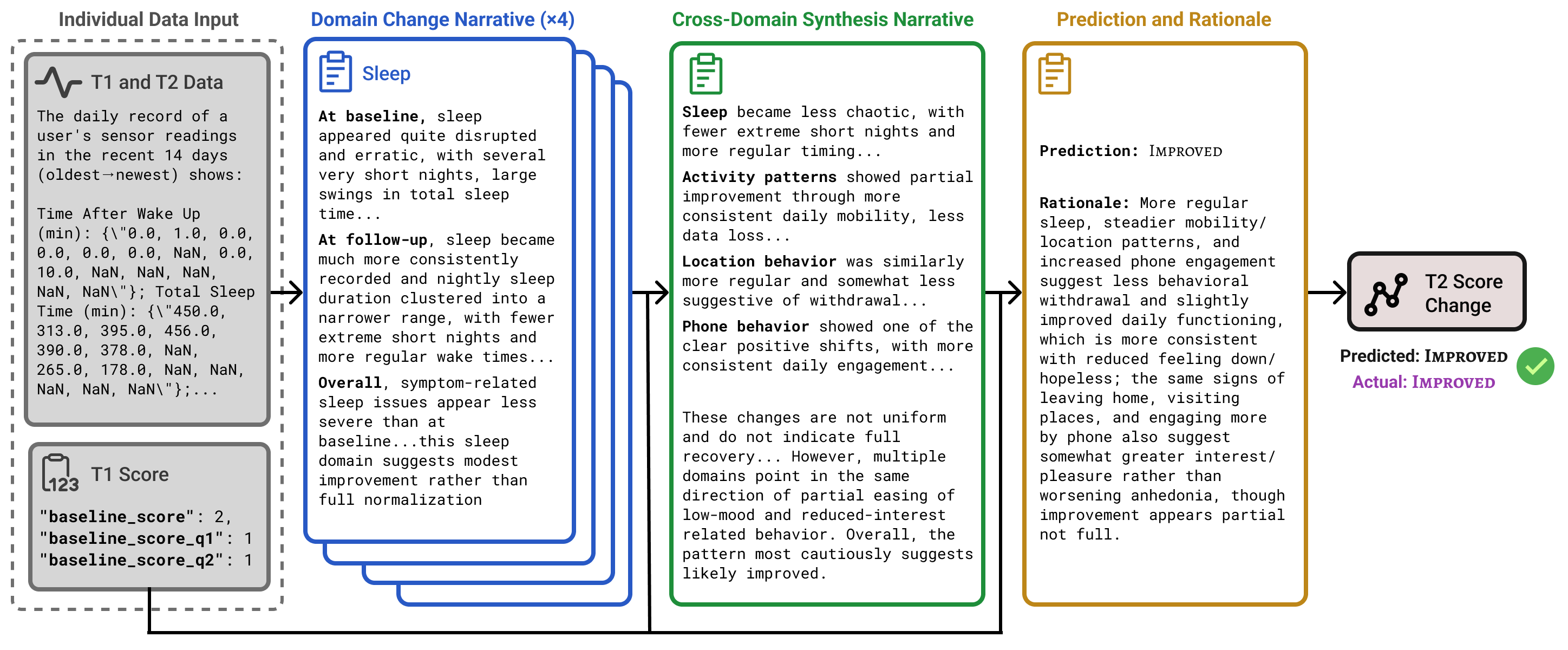}
    \caption{An illustrative OAK-Mind prediction for one participant in the GLOBEM depression task. The figure shows snippets from the sensing input and agent outputs.}
    \label{fig:instance}
\end{figure}

Figure~\ref{fig:instance} illustrates how OAK-Mind processed one participant's sensing data and baseline assessment. The example traces the information from a behavioral change narrative through cross-domain synthesis to the final prediction and rationale. The full set of original narratives for this instance is presented in Appendix \ref{app:instance}.

%% file: sections/04_evaluation.tex
\section{Prediction Task and Data Preparation}
\label{sec:task-data}

\subsection{Datasets and Behavioral Features}
We use two publicly available passive sensing and mental health datasets. Both provide longitudinal behavioral data and self-reported depression and anxiety measures through the PHQ-4.

\subsubsection{GLOBEM}
GLOBEM~\cite{xu2022globemdataset} is a multi-year passive sensing dataset containing wearable and mobile sensing data from 497 unique participants. It includes features related to location, phone usage, sleep, and steps, along with mental health measures collected through weekly ecological momentary assessments (EMA). 

\subsubsection{College Experience}
The College Experience Study Dataset~\cite{nepal2024capturing} (hereafter denoted as "College Experience") was collected through a four-year longitudinal mobile sensing study involving 215 first-year college students. It includes sensing data related to physical activity, mobility and semantic locations, audio plays, phone usage and sleep, alongside weekly self-reported measures of anxiety, depression, self-esteem, etc. 

\subsubsection{Behavioral Feature Selection}
To align the inputs with the Behavioral Change Agents, we manually selected features from each dataset's released feature set using two criteria: (1) the feature is conceptually related to at least one of the four behavioral domains (sleep, physical activity, mobility, and phone usage), and (2) its missing rate is below 30\% across participant-day records. This yielded 59 features for GLOBEM and 27 for College Experience. The complete feature lists are provided in Appendix~\ref{app:features}.

The feature sets vary across datasets. Since OAK-Mind reasons over natural-language descriptions rather than a fixed input schema, it does not require feature harmonization across datasets. All features are represented at the daily level, where features recorded at a higher frequency are averaged according to the date to obtain a single daily value.

\subsection{Prediction Instances and Data Splits}

\subsubsection{Participant Eligibility}
For GLOBEM, we use the last three cohorts as they contain complete PHQ-4 measures for each participant. And for College Experience, we restrict the sample to iOS users so that the same sensing features are available across participants, making it easier to be applied to supervised methods for comparison. From the records of this scoped participants, we extracted the selected features over the preceding 14 days and retained only responses with complete 14-day sensing data. We then identified participants with at least two eligible responses collected 50 or more days apart and randomly assigned them to disjoint training and test sets, so that no participant appears in both sets.

\subsubsection{Pair Construction}
Training instances used in supervised ML methods as well as test instances were constructed from pairs of eligible responses from the same participant. In the test set, each participant contributes exactly one pair with their earliest and latest eligible responses. This yields 124 test pairs for GLOBEM and 40 for College Experience.
Training participants may contribute multiple eligible pairs. To obtain a 70/30 ratio of training to test pairs, we randomly sampled 290 training pairs for GLOBEM and 93 for College Experience from all eligible pairs among training participants.

\section{Evaluation}
\label{sec:evaluation}

We evaluate OAK-Mind through three experiments examining (1) its performance relative to existing methods, (2) its robustness across pretrained LLM backbones, and (3) the contributions of its components.
We report accuracy, balanced accuracy, macro F1 score, and weighted F1 score for both depression and anxiety change prediction. Within each dataset, all methods are evaluated on the same test pairs.

\subsection{Baseline Comparisons}

We compare OAK-Mind with five supervised ML methods, four prior zero-shot LLM methods, and a majority-class baseline on GLOBEM and College Experience.

\subsubsection{Supervised ML Methods}
We selected supervised ML approaches adapted from prior sensing-based mental health studies. These methods cover regularized linear classification, personalized collaborative filtering, and tree-based ensembles, and also approaches that explicitly select features and combine information across sensing domains. The specific modeling approaches of the ML methods are outlined below:
\begin{itemize}
    \item Xu et al.~\cite{xu2021Collaborative} personalizes depression detection through memory-based collaborative filtering, weighting other participants' labels by behavioral relevance and aggregating feature-level predictions through majority voting.

    \item Saeb et al.~\cite{saeb_mobile_2015} classifies depressive symptoms using logistic regression with elastic-net regularization to limit overfitting when combining sensing features.

    \item Chikersal et al.~\cite{chikersalDetectingDepressionPredicting2021b}
    selects stable features using nested randomized logistic regression, trains logistic regression or gradient boosting models for individual sensing feature sets, and combines their prediction probabilities through an AdaBoost ensemble.

    \item Borelli et al.~\cite{Borelli2025Detection} combines missing-value imputation and SMOTE-based class balancing with LightGBM classification for multimodal passive sensing features.

    \item Ahmed et al.~\cite{Ahmed2025XGBoost} combines engineered actigraphy features with ADASYN-based class balancing and evaluates multiple classifiers for depression detection and severity classification. We adapt XGBoost, which is their best-performing classifier.
\end{itemize}

Given that the LLM methods receive behavioral data in ordered daily sequences without calendar dates, for each ML baseline, we derive method-specific representations from the 14 daily sensing readings preceding $T_1$ and $T_2$ according to the original implementations (Table~\ref{tab:ml_features}). Features requiring finer temporal resolution or weekday/weekend split are excluded because they cannot be reconstructed from the available daily readings used for LLM methods. The same feature-extraction rules are applied to training and test examples. The representations of both sensing windows together with the baseline subscale score constitute the model input. The target is the direction of score change from $T_1$ to $T_2$ as a 3-class classification (\textsc{Improved}, \textsc{Stable}, or \textsc{Worsened}).

\begin{table}[t]
\centering
\small
\renewcommand{\arraystretch}{1.1}
\caption{Sensing data representations used in our adaptations of existing ML and LLM methods.
All representations are derived separately for the 14-day windows preceding $T_1$ and $T_2$.}
\label{tab:ml_features}
\begin{tabular}{
    @{}>{\raggedright\arraybackslash}p{0.14\linewidth}
    p{0.25\linewidth}
    p{0.56\linewidth}@{}
}
\toprule
\textbf{Category} & \textbf{Method} & \textbf{Sensing data representation} \\
\midrule

ML Methods
& Xu et al.~\cite{xu2021Collaborative}
& Daily values categorized as low, medium, or high using the
33rd and 67th percentiles of the corresponding window. \\
\addlinespace

& Saeb et al.~\cite{saeb_mobile_2015}
& Mean of each sensing variable over the corresponding window. \\
\addlinespace

& Chikersal et al.~\cite{chikersalDetectingDepressionPredicting2021b}
& Mean and linear slope of each sensing variable over the full
14 days, the earliest seven days, and the latest seven days
in the corresponding window. \\
\addlinespace

& Borelli et al.~\cite{Borelli2025Detection}
& Mean, standard deviation, minimum, maximum, linear slope,
and proportion of missing days for each sensing variable
in the corresponding window. \\
\addlinespace

& Ahmed et al.~\cite{Ahmed2025XGBoost}
& Mean, standard deviation, interquartile range (IQR), skewness,
excess kurtosis, and lag-one autocorrelation for each sensing
variable in the corresponding window. \\
\midrule

Prior Zero-shot \ \ \ LLM Methods
& Kim et al.~\cite{kimHealthLLMLargeLanguage2024c}
& Daily average sensing sequences in each corresponding window. \\
\addlinespace

& Yu et al. (Zero-shot)~\cite{Yu2026Hybridsense}
& Daily average sensing sequences with skewness of daily mean step counts and minimum daily mean
sleep duration for the first and second 7 days in each corresponding window. \\
\addlinespace

& Yu et al. (Zero-shot with ToT)~\cite{Yu2026Hybridsense}
& Daily average sensing sequences with skewness of daily mean step counts and minimum daily mean
sleep duration for the first and second 7 days in each corresponding window.  \\
\addlinespace

& Yuan et al.~\cite{yuan2026digitalphenotyping}
& Daily average sensing sequences with the mean of each sensing variable over the 14-day corresponding window. \\
\midrule

Proposed Pipeline
& OAK-Mind
& Daily average sensing sequences in each corresponding window. \\

\bottomrule
\end{tabular}
\end{table}

\subsubsection{Zero-Shot LLM Methods}
We evaluate four prompting baselines adapted from three prior studies.
All are applied to the test set without training examples or
task-specific fine-tuning. Table~\ref{tab:ml_features} includes a summary of their adapted sensing inputs.
\begin{itemize}
    \item Kim et al.~\cite{kimHealthLLMLargeLanguage2024c} uses prompting framework that represents wearable sensor data as text and incorporates contextual information for health prediction. We adapt its zero-shot prompting approach.

    \item Yu et al. (Zero-shot)~\cite{Yu2026Hybridsense} uses structured prompts that specify the prediction target, sensing descriptors, health context, and output format to generate predictions without labeled examples.

    \item Yu et al. (Zero-shot with ToT)~\cite{Yu2026Hybridsense} uses the same sensing inputs with tree-of-thought prompting to enable LLMs exploring alternative reasoning paths and
    select the most plausible prediction.

    \item Yuan et al.~\cite{yuan2026digitalphenotyping} frames depression-change detection as a comparison between two two-week sensing windows. We adapt its zero-shot setting using daily
    sensing sequences and window-level feature means.
\end{itemize}

As with OAK-Mind, the zero-shot LLM baselines receive the baseline and follow-up sensing features ($X_{1,i}$ and $X_{2,i}$) in the form of chronological sequences. Together with the baseline subscale score $s_{1,i}$, the individual data are included in the prompts template demonstrated Table~\ref{tab:agents} to form the final input for all LLM methods.

\subsubsection{Majority-Class Baseline}
We include a majority-class baseline that assigns every test
instance the most frequent label in the corresponding training
set. This label is \textsc{Stable} for both prediction tasks
in both datasets.

\subsubsection{Implementation Details}
For the main comparison, all LLM methods, including OAK-Mind, use GPT-5.4~\cite{openai_gpt54} model with a temperature of 1.2. We run each LLM method five times and report the mean values for each of the performance metrics. Similarly, we also run each supervised ML method five times with different random seeds and report the mean performance.

\subsection{Generalizability Evaluation}

To examine whether OAK-Mind's performance gains extend across pretrained LLM backbones, we instantiate both OAK-Mind and the plain LLM baseline with each of four models: GPT-5.2~\cite{openai_gpt52}, Qwen-plus~\cite{alibaba_qwenplus}, Gemini-2.5-flash~\cite{google_gemini25flash}, and DeepSeek-chat~\cite{deepseek_api}. For each backbone, we evaluate both approaches on depression and anxiety change prediction using GLOBEM and report OAK-Mind's relative performance gains over the baseline using that same backbone.
The Plain LLM baseline receives the same baseline and follow-up sensing features with baseline subscale score as OAK-Mind. It then directly predicts the direction of mental health change from these inputs in a single prompt, without OAK-Mind’s domain-knowledge grounding, behavioral analysis, cross-domain synthesis, or questionnaire-based symptom rating.

\subsection{Ablation Study}
We conduct an ablation study using GPT-5.4~\cite{openai_gpt54} to examine the contributions of OAK-Mind’s components. We compare the full pipeline with four variants:

\begin{itemize}
    \item Without Domain Knowledge: removes the literature-derived domain knowledge provided to the behavioral agents while retaining the remaining pipeline.
    \item Unified Behavioral Agent: replaces the four domain-specific behavioral agents with one agent that jointly analyzes all behavioral domains using the same domain knowledge and general task instructions. Cross-domain synthesis and symptom rating are retained.
    \item Behavioral Analysis and Synthesis Only: retains the behavioral agents (\textcircled{1} in Figure\ref{fig:pipline}) and cross-domain synthesis (\textcircled{2} in Figure\ref{fig:pipline}) but removes the questionnaire-based Symptom Rating Agent (\textcircled{3} in Figure\ref{fig:pipline}).
    \item Questionnaire Agent Only: removes behavioral analysis and cross-domain synthesis (\textcircled{1} and \textcircled{2} in Figure\ref{fig:pipline}), providing the sensing features and baseline subscale score directly to the Symptom Rating Agent (\textcircled{3} in Figure\ref{fig:pipline}).
\end{itemize}

We also include the Plain LLM baseline described in Section 5.2 with GPT-5.4~\cite{openai_gpt54}, as a reference without any OAK-Mind components. To illustrate differences in how domain-specific and unified behavioral analysis organize and integrate evidence, we additionally compare their generated narratives for one GLOBEM depression instance with differing predictions.

%% file: sections/05_results.tex
\section{Results}
\label{sec:results}

In this section, we show that OAK-Mind, while without task-specific training, achieved the better or competitive performance on all reported metrics among the evaluated supervised ML and zero-shot LLM methods across both datasets and tasks. We first present these comparisons, then highlight that its generalizable advantage over direct prompting extends across all four tested LLM backbones on GLOBEM. Finally, ablation results indicate the complementary contributions of behavioral analysis and questionnaire-based symptom rating, as well as the benefits of domain-specific agents and domain knowledge components.

% ========== Modified Table 1 (Combined tasks) ==========

\begin{table*}[t]
\centering
\caption{Model performance comparison for depression and anxiety change classification on two datasets across traditional machine learning baselines, majority voting, and LLM-based methods. The best overall results are in \textbf{bold}, and the second best results are \underline{underlined}}
\label{tab:main_results}
\scriptsize
\setlength{\tabcolsep}{2pt}
\renewcommand{\arraystretch}{1}
\begin{tabular}{ll*{8}{C{1.2cm}}}
\toprule
\multirow{2}{*}{\textbf{Category}}
& \multirow{2}{*}{\textbf{Model}}
& \multicolumn{4}{c}{\textbf{GLOBEM}}
& \multicolumn{4}{c}{\textbf{College Experience}} \\
\cmidrule(lr){3-6} \cmidrule(lr){7-10}
&
& \textbf{Accuracy$\uparrow$}
& \textbf{Bal. Acc.$\uparrow$}
& \textbf{Macro F1$\uparrow$}
& \textbf{Wt. F1$\uparrow$}
& \textbf{Accuracy$\uparrow$}
& \textbf{Bal. Acc.$\uparrow$}
& \textbf{Macro F1$\uparrow$}
& \textbf{Wt. F1$\uparrow$} \\
\midrule

\multicolumn{10}{c}{\textit{Task: Depression Change} (\textsc{Improved}, \textsc{Stable} or \textsc{Worsened})} \\
\midrule
\multirow{1}{*}{Naive Baseline}
& Majority-Class Baseline & 0.45 & 0.33 & 0.21 & 0.29 & 0.40 & 0.33 & 0.19 & 0.23 \\
\addlinespace[4pt]

\multirow{5}{*}{\makecell{ML Methods}}
& Xu et al.~\cite{xu2021Collaborative} & 0.32 & 0.28 & 0.29 & 0.33 & 0.31 & 0.30 & 0.29 & 0.29 \\
& Saeb et al.~\cite{saeb_mobile_2015} & 0.39 & 0.39 & 0.37 & 0.38 & 0.38 & 0.35 & 0.33 & 0.35 \\
& Chikersal et al.~\cite{chikersalDetectingDepressionPredicting2021b} & 0.44 & 0.43 & 0.40 & 0.40 & \underline{0.48} & \underline{0.44} & 0.41 & \underline{0.44} \\
& Borelli et al.~\cite{Borelli2025Detection} & 0.43 & 0.40 & \underline{0.44} & 0.40 & 0.47 & \underline{0.44} & \underline{0.44} & \underline{0.44} \\
& Ahmed et al.~\cite{Ahmed2025XGBoost} & \underline{0.47} & \underline{0.46} & 0.43 & \underline{0.44} & 0.38 & 0.34 & 0.33 & 0.35 \\
\addlinespace[4pt]

\multirow{4}{*}{\makecell{Prior Zero-shot \\ LLM Methods}}
& Kim et al.~\cite{kimHealthLLMLargeLanguage2024c} & 0.33 & 0.33 & 0.33 & 0.33 & 0.41 & \underline{0.44} & 0.40 & 0.40 \\
& Yu et al. (Zero-shot)~\cite{Yu2026Hybridsense} & 0.34 & 0.36 & 0.34 & 0.34 & 0.40 & 0.43 & 0.39 & 0.38 \\
& Yu et al. (With TOT)~\cite{Yu2026Hybridsense} & 0.31 & 0.31 & 0.30 & 0.31 & 0.34 & 0.37 & 0.33 & 0.32 \\
& Yuan et al.~\cite{yuan2026digitalphenotyping} & 0.29 & 0.30 & 0.29 & 0.29 & 0.39 & 0.42 & 0.38 & 0.37 \\
\addlinespace[4pt]

\multirow{1}{*}{\textbf{Proposed Pipeline}}
& \textbf{OAK-Mind}
& \textbf{0.56} & \textbf{0.57} & \textbf{0.46} & \textbf{0.48}
& \textbf{0.52} & \textbf{0.52} & \textbf{0.51} & \textbf{0.52} \\

\midrule
\multicolumn{10}{c}{\textit{Task: Anxiety Change} (\textsc{Improved}, \textsc{Stable} or \textsc{Worsened})} \\
\midrule
\multirow{1}{*}{Naive Baseline}
& Majority-Class Baseline & 0.42 & 0.33 & 0.20 & 0.25 & 0.35 & 0.33 & 0.19 & 0.23 \\
\addlinespace[4pt]

\multirow{5}{*}{\makecell{ML Methods}}
& Xu et al.~\cite{xu2021Collaborative} & 0.37 & 0.35 & 0.35 & 0.36 & 0.34 & 0.34 & 0.32 & 0.32 \\
& Saeb et al.~\cite{saeb_mobile_2015} & \underline{0.44} & \underline{0.42} & \textbf{0.42} & \underline{0.43} & \underline{0.47} & \underline{0.46} & \underline{0.45} & \underline{0.46} \\
& Chikersal et al.~\cite{chikersalDetectingDepressionPredicting2021b} & \underline{0.44} & \underline{0.42} & \textbf{0.42} & 0.42 & 0.43 & 0.43 & 0.41 & 0.41 \\
& Borelli et al.~\cite{Borelli2025Detection} & 0.40 & 0.37 & 0.37 & 0.39 & 0.45 & 0.45 & 0.42 & 0.42 \\
& Ahmed et al.~\cite{Ahmed2025XGBoost} & 0.42 & 0.40 & \underline{0.40} & 0.41 & 0.43 & 0.43 & 0.41 & 0.41 \\
\addlinespace[4pt]

\multirow{4}{*}{\makecell{Prior Zero-shot \\ LLM Methods}}
& Kim et al.~\cite{kimHealthLLMLargeLanguage2024c} & 0.37 & 0.34 & 0.34 & 0.36 & 0.33 & 0.33 & 0.32 & 0.32 \\
& Yu et al.(Zero-shot)~\cite{Yu2026Hybridsense} & 0.35 & 0.33 & 0.33 & 0.35 & 0.32 & 0.32 & 0.30 & 0.30 \\
& Yu et al. (With TOT)~\cite{Yu2026Hybridsense} & 0.37 & 0.34 & 0.34 & 0.37 & 0.30 & 0.30 & 0.26 & 0.26 \\
& Yuan et al.~\cite{yuan2026digitalphenotyping} & 0.38 & 0.35 & 0.35 & 0.37 & 0.33 & 0.33 & 0.31 & 0.31 \\
\addlinespace[4pt]

\multirow{1}{*}{\textbf{Proposed Pipeline\ \ }}
& \textbf{OAK-Mind}
& \textbf{0.45} & \textbf{0.43} & \textbf{0.42} & \textbf{0.44}
& \textbf{0.48} & \textbf{0.47} & \textbf{0.47} & \textbf{0.48} \\

\bottomrule
\end{tabular}

\vspace{2pt}
\begin{flushleft}
\scriptsize
\textit{Note.} Bal. Acc. = balanced accuracy; Wt. F1 = weighted F1.
ML baselines are evaluated using a 70/30 train-test split.
\end{flushleft}
\end{table*}

\subsection{OAK-Mind Achieves Overall Best Performance Across Tasks and Datasets}

As shown in Table~\ref{tab:main_results}, OAK-Mind achieved the highest or joint-highest mean performance across all reported metrics for both tasks and datasets. Its advantage over the strongest supervised ML baselines was more pronounced for depression change classification. On GLOBEM, accuracy increased from 0.47 to 0.56 and balanced accuracy from 0.46 to 0.57; on College Experience, the corresponding improvements were from 0.48 to 0.52 and from 0.44 to 0.52. For anxiety change classification, the gains were smaller, but OAK-Mind matched or exceeded the best-performing ML baseline on each metric.

Across both datasets and tasks, supervised ML methods generally outperformed the prior zero-shot LLM methods, which makes them the strongest baselines in our evaluation. 

Taken together, these results demonstrate that without task-specific training, OAK-Mind was able to achieve competitive performance in the evaluated setting. It matched or exceeded both existing zero-shot approaches and the strongest supervised baselines across two datasets and two tasks, with especially pronounced gains in depression change classification.

% % ========== Generalizability ==========
\begin{figure}
    \centering
    \includegraphics[width=1.0\linewidth]{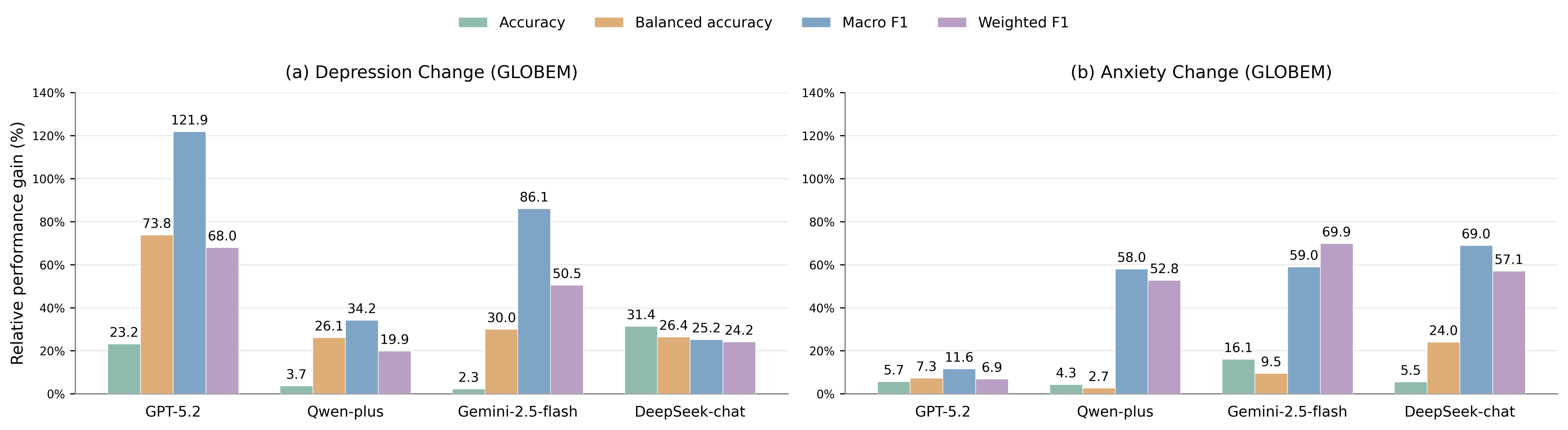}
    \caption{Relative performance gains of OAK-Mind over direct prompting on GLOBEM (\%) on (a) depression change and (b) anxiety change prediction. For each backbone, OAK-Mind and the plain LLM baseline use the same pretrained model.}
    \label{fig:generalizability}
\end{figure}

\subsection{OAK-Mind's Robust Gains Across LLM Backbones}

Figure~\ref{fig:generalizability} indicates that OAK-Mind improved the performance from LLM baseline with direct prompting on every reported metric for both tasks across all four LLM backbones tested with GLOBEM. The magnitude of improvement varied by backbone and task. For depression, GPT-5.2 showed the largest relative gains in balanced accuracy and macro F1 (73.8\% and 121.9\%, respectively). For anxiety, DeepSeek-Chat showed the largest gains on these metrics (24.0\% and 69.0\%). These results show a consistent benefit over direct prompting across the tested backbones.

% % ========== Ablation ==========
\begin{table*}[t]
\centering
\caption{Ablation study for depression and anxiety change classification
on two datasets. The best results for each dataset and task are
in \textbf{bold}.}
\label{tab:ablation_results}
\scriptsize
\setlength{\tabcolsep}{2pt}
\renewcommand{\arraystretch}{1.15}

\begin{tabular}{lccc*{9}{C{0.98cm}}}
\toprule
\multirow{2}{*}{\textbf{Variant}}
& \multirow{2}{*}{\makecell{\textbf{Behavioral}\\\textbf{Analysis \&}\\\textbf{Synthesis}}}
& \multirow{2}{*}{\makecell{\textbf{Domain}\\\textbf{Knowledge}}}
% & \multirow{2}{*}{\makecell{\textbf{Behavioral}\\\textbf{Analysis \&}\\\textbf{Synthesis}}}
& \multirow{2}{*}{\makecell{\textbf{Domain}\\\textbf{-Specific}\\\textbf{Agents}}}
& \multirow{2}{*}{\makecell{\textbf{PHQ-4}\\\textbf{Filling}}}
& \multicolumn{4}{c}{\textbf{GLOBEM}}
& \multicolumn{4}{c}{\textbf{College Experience}} \\
\cmidrule(lr){6-9} \cmidrule(lr){10-13}
& & & &
& \textbf{Accuracy$\uparrow$} 
& \textbf{Bal. Acc.$\uparrow$} 
& \textbf{Macro F1$\uparrow$} 
& \textbf{Wt. F1$\uparrow$}
& \textbf{Accuracy$\uparrow$} 
& \textbf{Bal. Acc.$\uparrow$} 
& \textbf{Macro F1$\uparrow$} 
& \textbf{Wt. F1$\uparrow$} \\
\midrule

\multicolumn{13}{c}{\textit{Task: Depression Change}} \\
\midrule

Full pipeline
& \checkmark & \checkmark & \checkmark & \checkmark
& \textbf{0.56} & \textbf{0.57} & 0.46 & \textbf{0.48}
& \textbf{0.52} & \textbf{0.52} & \textbf{0.51} & \textbf{0.52} \\

Without Domain Knowledge
&\checkmark & & \checkmark & \checkmark
& 0.53 & 0.54 & \textbf{0.48} & \textbf{0.48}
& 0.49 & 0.51 & 0.49 & 0.50 \\

Unified Behavioral Agent
& \checkmark & \checkmark & & \checkmark
& 0.46 & 0.49 & 0.40 & 0.41
& 0.50 & 0.51 & 0.50 & 0.51 \\

Behavioral and Synthesize Only
& \checkmark & \checkmark & \checkmark &
& 0.37 & 0.43 & 0.32 & 0.33
& 0.43 & 0.44 & 0.42 & 0.41 \\

Questionnaire Agent Only
& & & & \checkmark
& 0.46 & 0.36 & 0.27 & 0.32
& 0.39 & 0.41 & 0.37 & 0.37 \\

Plain LLM Baseline
& & & &
& 0.45 & 0.33 & 0.21 & 0.29
& 0.40 & 0.33 & 0.19 & 0.23 \\

\midrule
\multicolumn{13}{c}{\textit{Task: Anxiety Change}} \\
\midrule

Full pipeline
& \checkmark & \checkmark & \checkmark & \checkmark
& \textbf{0.45} & \textbf{0.42} & \textbf{0.42} & \textbf{0.44}
& \textbf{0.48} & \textbf{0.47} & \textbf{0.47} & \textbf{0.48} \\

Without Domain Knowledge
& & \checkmark & \checkmark & \checkmark
& 0.44 & 0.41 & 0.41 & 0.41
& 0.44 & 0.44 & 0.44 & 0.44 \\

Unified Behavioral Agent
& \checkmark & \checkmark & & \checkmark
& 0.40 & 0.39 & 0.39 & 0.40
& 0.40 & 0.40 & 0.40 & 0.40 \\

Behavioral and Synthesize Only
& \checkmark & \checkmark & \checkmark &
& 0.44 & 0.40 & 0.40 & 0.43
& 0.37 & 0.36 & 0.33 & 0.34 \\

Questionnaire Agent Only
& & & & \checkmark
& 0.40 & 0.35 & 0.34 & 0.37
& 0.31 & 0.32 & 0.26 & 0.26 \\

Plain LLM Baseline
& & & &
& 0.42 & 0.33 & 0.20 & 0.25
& 0.33 & 0.34 & 0.30 & 0.29 \\

\bottomrule
\end{tabular}

\vspace{3pt}
\begin{minipage}{\textwidth}
\scriptsize
\textit{Note.}
Bal. Acc. = balanced accuracy; Wt. F1 = weighted F1. \\
\textit{Domain-Specific Agents} indicates that behavioral domains are analyzed by separate agents; the unified variant (\textit{Unified behavioral agent}) analyzes all domains jointly using the same domain knowledge.
\end{minipage}
\end{table*}

\subsection{Contributions of Pipeline Components}
As shown in Table~\ref{tab:ablation_results}, the OAK-Mind pipeline, or the combination of knowledge-grounded behavioral analysis and synthesis with questionnaire agent outperformed either component alone across both tasks and datasets. Even without domain knowledge, adding domain-specific behavioral analysis and synthesis to the Questionnaire Agent Only condition improved every metric; especially for depression on GLOBEM, where macro F1 increased from 0.27 to 0.48. Similarly, including questionnaire agent to enable PHQ-4 filling after knowledge-grounded behavioral analysis and synthesis improved every metric compared with Behavioral and Synthesize Only condition. For anxiety, this increased macro F1 from 0.40 to 0.42 on GLOBEM and from 0.33 to 0.47 on College Experience.

Domain knowledge provided modest additional gains by increasing accuracy and balanced accuracy by 0.01 to 0.04 across all four dataset-task combinations. Both F1 measures improved in most of the cases. The exception was depression change detection on GLOBEM, where macro F1 decreased from 0.48 to 0.46 and weighted F1 remained unchanged at 0.48.

\textbf{Contribution of domain-specific agents.}
With the same domain knowledge and general task instructions, the full pipeline outperformed the variant using a unified behavioral agent on all reported metrics across both datasets and tasks (Table~\ref{tab:ablation_results}). 
For depression, domain-specific agents increased balanced accuracy and macro F1 by 0.08 and 0.06 on GLOBEM, respectively, compared with 0.01 on both metrics on College Experience. For anxiety, both metrics increased by 0.04 on GLOBEM and 0.07 on College Experience. 
This comparison suggests that explicit domain-level analysis before synthesis improves inference beyond supplying the same knowledge to a single behavioral agent.

\begin{figure}
    \centering
    \includegraphics[width=1\linewidth]{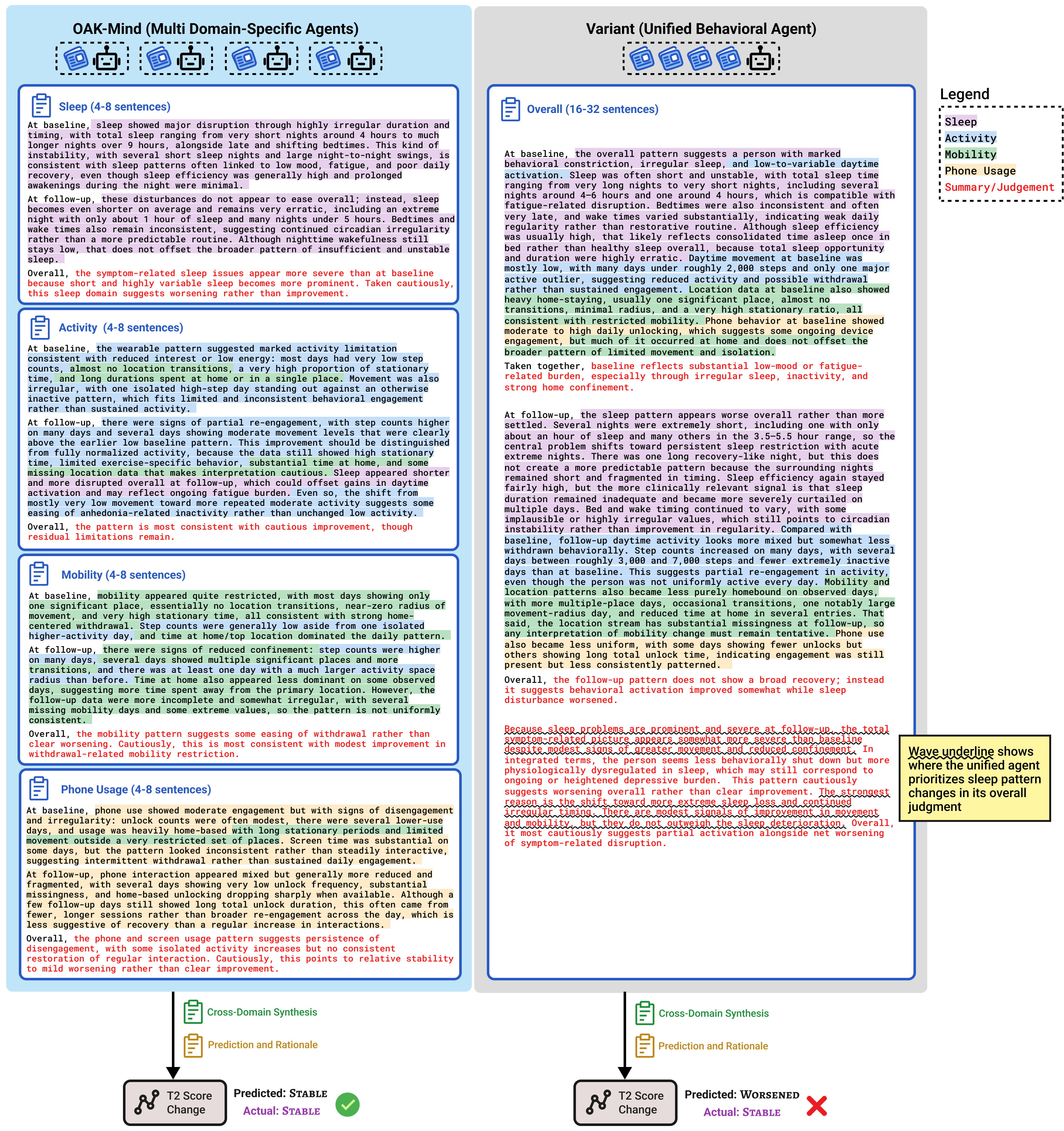}
    \caption{An illustrative comparison of domain-specific and unified behavioral analysis for a GLOBEM depression instance. OAK-Mind produces separate domain-level assessments before cross-domain synthesis, whereas the unified variant interprets all domains within one narrative. \hl{Highlight colors} identify the behavioral domains discussed, \textcolor{red}{red text} marks summary judgments, and \uwave{wave underlines} identify statements that prioritize sleep changes in the unified interpretation. With synthesis and symptom rating agents, OAK-Mind predicts \textsc{Stable} correctly for this case while the unified variant predicts \textsc{Worsened}.}
    \label{fig:casestudy}
\end{figure}

To illustrate how domain-specific decomposition can shape behavioral interpretation, we present a case study in Figure~\ref{fig:casestudy} that compares the generated behavioral change narratives for one GLOBEM depression instance with an unchanged subscale score of zero. According to the narratives, both configurations identify worsening sleep alongside partial improvement in activity and mobility, while phone engagement remains irregular. OAK-Mind articulates these changes through separate domain-level assessments: the sleep agent identifies worsening, the activity and mobility agents identify cautious improvement, and the phone agent identifies stability to mild worsening. While individual agents also reference related patterns from other domains as they have access to the full feature input, each provides an explicit assessment centered on its assigned behavioral domain.

These separate assessments provide the subsequent synthesis stage with distinct, potentially conflicting interpretations. In this case, OAK-Mind’s synthesis retains the mixed directions of change, and the final symptom rating correctly predicts stability. The unified agent also recognizes increased activity and reduced confinement, but its integrated judgment gives greater weight to sleep pattern changes as it states the daytime gains "do not outweigh the sleep deterioration", and ultimately wrongly predicted the change to be \textsc{Worsened}. 
This selected case illustrates a potential benefit of maintaining explicit domain-level assessments before synthesis when behavioral changes point in different directions.

%% file: sections/06_discussion.tex
\section{Discussion}

\subsection{Toward Longitudinal Inference Without Task-Specific Training}
A central challenge in longitudinal mental health sensing is that supervised models often depend on labeled examples linking behavioral data to repeated assessments. In our work, OAK-Mind explores an alternative starting point: a person’s own baseline assessment and sensing data serve as a reference for interpreting later behavioral patterns, without task-specific model training. This formulation allows later change to be examined relative to the individual’s earlier patterns without requiring a participant-specific model.

In the scenario considered here, a person completes a mental health assessment and contributes sensing data from their smartphone and wearable. At a later point, OAK-Mind compares a new sensing window with the baseline to infer whether self-reported symptoms may have improved, remained stable, or worsened. In our evaluation, the pipeline achieved higher performance across the reported metrics than the tested supervised baselines, despite not using their task-specific training pairs. These results suggest that knowledge-grounded LLMs can offer an approach to longitudinal inference when labeled data for the target task are limited.

This formulation also separates the timing of inference from the timing of self-report. Once a baseline is available, the system can generate an estimate from a later sensing window without requiring a new questionnaire at that point. In the longer term, this approach could create additional opportunities to identify potential symptom changes between scheduled self-report, providing a complementary reference alongside those assessments.

\subsection{Connecting Behavioral Change with Self-Reported Symptoms}
Passive sensing and self-report questionnaires offer different views of mental health: sensors capture patterns of behavior, while questionnaires ask people to report their symptoms. OAK-Mind brings these views together by organizing inference from domain-specific behavioral changes to questionnaire-based symptom assessment. In the ablation study, combining domain-specific behavioral analysis and synthesis with symptom rating outperformed either tested configuration alone across both datasets and tasks. This finding points to the value of making the relationship between sensed behavior and the target symptom measure an explicit part of the inference task.

Domain knowledge grounding provides a further scaffold for this process. In our system, each behavioral agent receives literature-derived associations relevant to its domain and the target outcome, guiding its interpretation of changes in the sensing data. Adding this knowledge produced modest gains in accuracy and balanced accuracy across both tasks and datasets, although improvements were not consistent across all metrics. 

Together, these results suggest that combining behavioral analysis with symptom-oriented assessment can provide a useful structure for training-free inference, with domain knowledge offering a further scaffold for interpreting behavioral change.

\subsection{Making Intermediate Mental Health Inference Traceable}
OAK-Mind is designed to produce narratives at each stage of inference, making its intermediate outputs available for inspection. As illustrated in Figure~\ref{fig:instance}, a behavioral agent describes changes within its assigned domain, the synthesis agent brings together the domain-level analyses, and the symptom rating agent relates the synthesis to questionnaire items. These outputs allow a reader to follow the pipeline's stated interpretation from behavioral data to predicted symptom change.

The narratives provide opportunities to examine how a behavioral pattern is described, how evidence across domains is combined, and how the resulting synthesis is related to symptoms. They may also help identify where an interpretation needs additional context. For example, an agent may accurately describe reduced mobility, while the individual could explain that it reflects an injury or a change in schedule.

Such intermediate outputs offer a starting point for examining mental health inferences from passive sensing. Future systems could build on this structure to let individuals or clinicians review and contextualize predictions. Our evaluation focused on classification performance; further work is required to examine the accuracy and faithfulness of the narratives and if they help people audit the pipeline's outputs.

\subsection{Limitations and Future Work}
While OAK-Mind demonstrates the potential of training-free inference of mental health change from passive sensing data, several limitations should be acknowledged.

\textit{Mental health measures.} In this work, we defined change using differences in PHQ-2 and GAD-2 scores and treating any increase or decrease as a change in symptoms. These brief measures provide a consistent reference for evaluation, but a change in self-reported score does not necessarily indicate a clinically meaningful change. Our evaluation did not include clinical assessments or clinician judgments; the results therefore reflect agreement with changes in self-reported symptoms rather than clinical diagnoses or judgments. 

\textit{Population and sensing conditions.}
Both datasets consist of college students, and our analysis of College Experience was limited to iOS users. We also retained only assessments with complete 14-day sensing windows. These choices leave open how OAK-Mind would perform with other populations, devices, or less complete sensing data. 
Moreover, the domain knowledge used by the agents reflects associations reported in prior literature, which may not apply equally across individuals and circumstances. Future work could examine performance across more diverse populations and investigate whether incorporating knowledge relevant to a person’s context improves inference.

\textit{Inference and use in practice.}
Our evaluation examined classification performance, but did not assess whether the intermediate narratives accurately describe the sensing data or provide well-grounded interpretations of symptoms. Nor did it test how people might use these narratives to review a prediction. Future work could evaluate the narratives against the underlying data and with relevant experts, as well as study how individuals interpret predictions and contribute context.

%% file: sections/07_conclusion.tex
\section{Conclusion}

We presented OAK-Mind, a training-free, knowledge-grounded multi-agent LLM pipeline for inferring changes in depression and anxiety symptoms from passive sensing data. OAK-Mind organizes inference into three steps, behavioral analysis, cross-domain synthesis, and questionnaire-based symptom rating, using domain knowledge to guide the interpretation of behavioral changes.
Evaluations across two datasets and both tasks demonstrated that OAK-Mind achieved higher performance across the reported metrics than the tested supervised and zero-shot LLM baselines. It also showed consistent gains over direct prompting on four LLM backbones across all metrics. Ablation results validated the benefits brought by components including behavioral analysis and synthesis, questionnaire-based symptom rating, domain knowledge grounding and domain-specific agent design.
These findings highlight the potential of structured LLM-based inference for longitudinal mental health research when task-specific labeled data are limited. By comparing later sensing data with an individual’s baseline and producing intermediate narratives, OAK-Mind offers a way to examine how behavioral changes may relate to symptom predictions, as well as provides a starting point for more personalized, interpretable, and label-efficient mental health inference from passive sensing.

%% file: sections/appx_kg.tex
\section{Knowledge Graph Construction}
\label{app:kg}

\textbf{Sources and scope.} The construction corpus of this mental health knowledge graph included research literature retrieved from PubMed, APA PsycInfo, CINAHL Complete, IEEE Xplore, and the ACM Digital Library, supplemented by clinical guidelines and related reference documents. The literature searches combined terms describing sensing modalities and behavioral signals, mental-health conditions, and clinical processes such as screening, monitoring, assessment, and intervention, detailed as below.

Two complementary Boolean queries shared the same sensing and
mental-health blocks and differed in their clinical-process terms:

\begin{quote}
\textbf{Query 1:} Sensing AND Detection AND Mental health\\
\textbf{Query 2:} Sensing AND Intervention AND Mental health
\end{quote}

Within each block, terms were combined using OR.

\paragraph{Sensing block: shared by both queries.}
\begin{quote}
\small\ttfamily\raggedright
("digital phenotype*" OR "sensor" OR "behavioral pattern"
OR "wearable" OR "smartphone" OR "screen use" OR "screen usage"
OR "location" OR "GPS" OR "digital biomarkers" OR "digital health"
OR "smartwatch" OR "mobile health"
OR "ecological momentary assessment" OR "heart rate"
OR "accelerometer" OR "IMU" OR "self-report behavior"
OR "sleep duration" OR "sleep stage" OR "sleep pattern"
OR "sleep quality" OR "behavioral symptom"
OR "physical activity intensity" OR "physical activity pattern"
OR "fitness tracker" OR "mobility pattern" OR "daily routine"
OR "biofeedback" OR "skin conductance" OR "EDA"
OR "galvanic skin response" OR "PPG" OR "photoplethysmography"
OR "EEG" OR "electroencephalography" OR "voice analysis"
OR "speech analysis" OR "acoustic features" OR "vocal biomarker"
OR "typing pattern" OR "keystroke")
\end{quote}

\paragraph{Detection block: Query 1.}
\begin{quote}
\small\ttfamily\raggedright
("diagnosis" OR "screening" OR "monitoring"
OR "prediction" OR "detection" OR "classification")
\end{quote}

\paragraph{Intervention block: Query 2.}
\begin{quote}
\small\ttfamily\raggedright
("assessment" OR "treatment" OR "intervention"
OR "outcome" OR "effectiveness" OR "measurement" OR "evaluation")
\end{quote}

\paragraph{Mental-health block: shared by both queries.}
\begin{quote}
\small\ttfamily\raggedright
("Mental disorder" OR "Mental Health" OR "Dysthymia"
OR "Persistent Depressive Disorder" OR "GAD"
OR "Generalized Anxiety Disorder" OR "Panic Disorder"
OR "Agoraphobia" OR "Social Phobia" OR "Social Anxiety Disorder"
OR "Specific Phobia" OR "Separation Anxiety Disorder"
OR "PTSD" OR "Post-Traumatic Stress Disorder"
OR "Acute Stress Disorder" OR "OCD"
OR "Obsessive-Compulsive Disorder" OR "Body Dysmorphic Disorder"
OR "Schizophrenia Spectrum" OR "Schizophrenia"
OR "Mood Disorders with Psychosis" OR "Bipolar I Disorder"
OR "Bipolar II Disorder" OR "Bipolar Disorder"
OR "Schizotypal PD" OR "Schizotypal Personality Disorder"
OR "Antisocial PD" OR "Antisocial Personality Disorder"
OR "Conduct Disorder" OR "ODD" OR "Oppositional Defiant Disorder"
OR "IED" OR "Intermittent Explosive Disorder"
OR "Narcissistic PD" OR "Narcissistic Personality Disorder"
OR "Antisocial PD" OR "Antisocial Personality Disorder"
OR "Borderline PD" OR "Borderline Personality Disorder"
OR "Paranoid PD" OR "Paranoid Personality Disorder"
OR "Schizoid PD" OR "Schizoid Personality Disorder"
OR "Avoidant PD" OR "Avoidant Personality Disorder"
OR "Histrionic PD" OR "Histrionic Personality Disorder"
OR "Somatic Symptom Disorder" OR "Illness Anxiety Disorder"
OR "Anorexia Nervosa" OR "Bulimia Nervosa"
OR "Binge-Eating Disorder" OR "Eating Disorder"
OR "Low Desire" OR "Arousal Difficulties"
OR "Orgasmic Dysfunction" OR "Sexual Pain"
OR "ADHD" OR "Attention-Deficit/Hyperactivity Disorder"
OR "Inattention" OR "Hyperactivity" OR "Impulsivity"
OR "ASD" OR "Autism Spectrum Disorder" OR "Autistic"
OR "Substance Use Disorder" OR "Alcohol Use Disorder"
OR "Insomnia" OR "Schizoaffective Disorder"
OR "Adjustment Disorder" OR "Dependent Personality Disorder"
OR "Obsessive-Compulsive Personality Disorder"
OR "Depressive Disorder" OR "suicidal ideation" OR "self-harm")
\end{quote}

Literature searches were conducted on 22 June 2026.
The search yielded 73,225 unique records. We used

\textbf{LLM-based relevance screening.} The abstracts of the literature underwent LLM-based relevance screening. GPT-5 mini assessed each unit on three criteria, each scored from 0 to 3: mental-health relevance, mobile/wearable-sensing relevance, and clinical actionability. Each unit was evaluated in three separate runs, and scores were averaged across runs. Units were retained if their mean mental-health relevance score was at least 2 and either their mean sensing-relevance or clinical-actionability score was at least 2. This procedure assessed relevance to the graph's scope, rather than the methodological quality or clinical correctness of individual claims. 48,427 literature items passed the LLM-based screen.

\textbf{Construction process.}
Following LLM-based relevance screening and text preprocessing,
entities and relationships were extracted from the source material using Gemini 2.5 Flash. The extracted records underwent
normalization, entity resolution, and relationship deduplication
before being assembled into the knowledge graph. The graph contained 1,223,018 entity records and 3,116,069 relationship records.

%% file: sections/appx_prompt.tex
\section{OAK-Mind Prompts for Depression Change Prediction on GLOBEM}
\label{app:prompt}

The following prompts reproduce the GLOBEM depression-change configuration. The original JSONL prompt files for all experimental configurations are provided in the supplementary materials.

\subsection{Behavioral Change Agents}
\subsubsection{Sleep change agent}

You are a sleep behavior expert analyzing wearable-derived sleep features and their association with depression.
  
  DOMAIN KNOWLEDGE: Improvement in depression-related sleep does not require optimal sleep. It often appears as easing of disruptions, reduced variability, fewer extreme nights, or more predictable timing.
  
  BASELINE (Time 1): \{baseline features text\}, FOLLOW-UP (Time 2):\{followup features text\}
  
  TASK: Compare baseline vs follow-up sleep data and describe how symptom-related sleep patterns evolve. In your 4–8 sentences:
  
  - Describe baseline sleep disruptions relevant to low mood or fatigue.
  
  - Describe whether follow-up sleep shows easing of these disruptions, persistence, or worsening.
  
  - Explicitly state if symptom-related sleep issues are less severe, similar, or more severe than baseline.
  
  - Treat partial easing as meaningful improvement.
  
  - End with one cautious sentence indicating whether this domain suggests improvement, stability, or worsening.
  
  - Third-person only.

\subsubsection{Activity change agent}
You are a physical activity expert analyzing wearable-derived activity features and their association with depression.

DOMAIN KNOWLEDGE: Improvement in anhedonia-related activity patterns often appears as reduced inactivity, more consistent daily movement, or partial re-engagement rather than high activity levels.

BASELINE (Time 1):\{baseline features text\}, FOLLOW-UP (Time 2):\{followup features text\}

TASK: Compare baseline vs follow-up activity data and describe symptom-related changes. In your 4–8 sentences:

- Describe baseline activity limitations linked to reduced interest or energy.

- Describe whether follow-up activity shows easing, persistence, or worsening of these limitations.

- Treat partial re-engagement as meaningful improvement.

- Explicitly distinguish improvement from unchanged low activity.

- End with one cautious sentence indicating improvement, stability, or worsening.

- Third-person only. 

\subsubsection{Location change agent}
You are a mobility and location behavior expert analyzing location-derived features and their association with depression.

DOMAIN KNOWLEDGE: Improvement may involve reduced withdrawal, slightly increased outings, or more consistent mobility patterns, even if overall movement remains limited.

BASELINE (Time 1): \{baseline features text\}, FOLLOW-UP (Time 2):\{followup features text\}

TASK: Compare baseline vs follow-up mobility data in terms of symptom-related withdrawal. In your 4–8 sentences:

- Describe baseline mobility restriction related to withdrawal.

- Describe whether follow-up shows easing of withdrawal, no change, or worsening.

- Treat reduced confinement or increased consistency as improvement.

- End with one cautious sentence indicating improvement, stability, or worsening.

- Third-person only.

\subsubsection{Phone change agent}
You are a smartphone usage behavior expert analyzing phone and screen usage features and their association with depression.

DOMAIN KNOWLEDGE: Easing of depressive symptoms may appear as reduced disengagement in phone usage, more regular phone use, or modest increases in interaction.

BASELINE (Time 1): \{baseline features text\}, FOLLOW-UP (Time 2):\{followup features text\}

TASK: Compare baseline vs follow-up phone and screen usage and describe symptom-related changes. In your 4–8 sentences:

- Describe baseline disengagement in phone usage patterns.

- Describe whether follow-up phone interaction and usage shows easing, persistence, or worsening.

- Treat partial re-engagement in phone usage as meaningful improvement.

- End with one cautious sentence indicating improvement, stability, or worsening.

- Third-person only. 

\subsection{Cross-Domain Synthesis Agent}
You are synthesizing behavioral changes across multiple domains to assess depression state change.

INPUTS:
- Sleep:\{sleep change narrative\}
- Activity:\{activity change narrative\}
- Location:\{location change narrative\}
- Phone:\{phone change narrative\}

Baseline depression score at Time 1:\{baseline score\}

TASK: Write 6–8 sentences that:

- Assess whether symptom-related signals (low mood, reduced interest) have eased, persisted, or worsened.

- Treat easing of symptoms as improvement even if full resolution is not present.

- Identify whether multiple domains show consistent easing.

- End with a cautious conclusion: likely improved, likely stable, or likely worsened.

- Third-person only.

\subsection{Symptom Rating Agent}
Based on the following emotional change assessment derived from behavioral sensor data:\{emotional change narrative\}:

The PHQ-4 depression subscale assesses depression through two core symptoms: 1. Feeling down, depressed, or hopeless; 2. Little interest or pleasure in doing things. Your task is to predict how this person's depression has CHANGED from Time 1 to Time 2.

BASELINE INFORMATION (Time 1): The scale is: 0 = Not at all, 1 = Several days, 2 = More than half the days, 3 = Nearly every day. 

- Question a) Feeling down, depressed or hopeless: \{baseline score q1\} out of 3. 

- Question b) Little interest or pleasure in doing things: \{baseline score q2\} out of 3.

Analyze each symptom dimension based on the behavioral changes:

- How do the observed behavioral changes relate to feelings of depression or hopelessness?

- How do the observed behavioral changes relate to interest or pleasure in activities?

Then make an overall judgment:

- If behavioral patterns suggest IMPROVEMENT in depressive symptoms → Improved (-1)

- If behavioral patterns suggest WORSENING of depressive symptoms → Worsened (+1)

- If behavioral patterns suggest NO CLEAR CHANGE or MIXED signals → Stable (0)

Output exactly one line in the following format: <predicted change> ||| <brief reason explaining the change, referencing both symptom dimensions>, where <predicted change> is: -1 (improved), 0 (stable), or 1 (worsened). Do not include any extra text.

%% file: sections/appx_instance.tex
\section{Output narratives of OAK-Mind for the instance in Figure \ref{fig:instance}}
\label{app:instance}

Below is the original narratives generated by OAK-Mind when preforming a depression change prediction (INS-W\_344 from GLOBEM dataset).

\textbf{Sleep Change Narrative: }
At baseline, sleep appeared quite disrupted and erratic, with several very short nights, large swings in total sleep time, and very late bedtimes on some recorded nights, all of which can relate to low mood, fatigue, and circadian instability. There were also many missing sleep days, which limits certainty but still suggests an inconsistent sleep pattern rather than a stable routine. At follow-up, sleep became much more consistently recorded and nightly sleep duration clustered into a narrower range, with fewer extreme short nights and more regular wake times. Although total sleep time still often remained on the short side and sleep efficiency weakened on some later nights, the overall pattern looks less chaotic and more predictable than before. Nighttime awakenings and time after final wake stayed generally low, so the main shift was improved regularity rather than fully restored or optimal sleep. Overall, symptom-related sleep issues appear less severe than at baseline, mainly because variability and extreme disruption decreased even though some insufficient sleep persists. Taken cautiously, this sleep domain suggests modest improvement rather than full normalization.

\textbf{Activity Change Narrative: }
At baseline, the activity pattern suggested reduced interest or energy through long stationary periods, high time spent at home or in one dominant location, and several days with very limited movement variety despite some isolated high-step days. The baseline record also had substantial missing data and increasingly irregular sleep toward the later days, which makes the picture less stable but still consistent with inactivity and disengagement rather than steady daily re-engagement. At follow-up, movement appeared more consistently recorded and more regular across days, with steps present every day and repeated evidence of leaving home, making multiple location transitions, and reaching several distinct places on at least some days. Although overall step counts were not uniformly higher and many days still showed high home time and substantial stationary behavior, the pattern was not simply unchanged low activity; it showed partial re-engagement through more consistent day-to-day mobility and less extreme data loss. Sleep also looked more consistently captured and somewhat more regular, which can fit with modest recovery in daily structure rather than a full return to high activity. The limitations therefore seemed eased in part, not resolved, because increased consistency and some broader mobility are meaningful improvements even without large gains in total activity. Overall, the sensor pattern cautiously suggests mild improvement rather than clear stability or worsening.

\textbf{Location Change Narrative: } 
At baseline, mobility already suggested only mild withdrawal rather than extreme confinement: step counts were often moderate to high, but several days showed strong concentration in one place, very high stationary time, and some zero-transition/one-place days that imply intermittent staying in. The baseline pattern also looked somewhat erratic, with large day-to-day swings in number of places visited, transitions, area covered, and time at home, which can reflect inconsistent engagement with the outside world. At follow-up, movement became more regular across the full two weeks, with complete data, steadier daily steps, and repeated days of moderate activity rather than sharp swings between highly mobile and highly restricted days. Although several follow-up days were still heavily home-centered and included one-place days, there were also multiple days with more transitions, more significant places, and sustained community movement, indicating that confinement was not pervasive. The reduced stationary ratio on a number of days and the more consistent mobility routine are more compatible with slightly less withdrawal than with worsening isolation. Overall, the pattern cautiously suggests mild improvement in withdrawal-related mobility behavior rather than clear deterioration.

\textbf{Phone Usage Change Narrative: }
At baseline, phone use looked intermittently disengaged and irregular: some days had very low unlock counts, delayed first phone use, and several stretches with missing data, suggesting inconsistent interaction. Even when the phone was used, behavior alternated between brief checking and a few unusually long sessions, which fits a pattern of uneven engagement rather than steady daily use. At follow-up, phone interaction became much more consistently present across days, with unlock activity recorded every day and generally higher unlock frequency than in the baseline period. Total unlock duration also increased on many days, with especially strong increases in home-based phone use, indicating more active and sustained interaction with the device. First unlock after midnight was often much earlier at follow-up, showing the phone was being engaged with sooner and more regularly rather than after long delays. Call behavior also appeared somewhat more active overall, particularly outgoing calls on several days, which is compatible with modest behavioral re-engagement.  Overall, the follow-up pattern suggests meaningful partial re-engagement in phone and screen use, cautiously consistent with improvement rather than worsening.

\textbf{Cross-Domain Synthesis narrative: }
Across sleep, activity, location, and phone behavior, several symptom-related signals appear to have eased rather than intensified. Sleep remained somewhat short and not fully normalized, but it became less chaotic, with fewer extreme short nights and more regular timing, suggesting modest improvement in circadian stability. Activity patterns also showed partial improvement through more consistent daily mobility, less data loss, and signs of leaving home and visiting multiple places on at least some days, even though overall activity was still not robust. Location behavior was similarly more regular and somewhat less suggestive of withdrawal, with fewer sharp swings and repeated evidence of community movement despite continued home-centered days. Phone behavior showed one of the clearer positive shifts, with more consistent daily engagement, higher unlock frequency, earlier use, and somewhat more calling, which is compatible with greater behavioral engagement. These changes are not uniform and do not indicate full recovery, since some reduced activity, home time, and short sleep still persisted. However, multiple domains point in the same direction of partial easing of low-mood and reduced-interest related behavior. Overall, the pattern most cautiously suggests likely improved.

\textbf{Questionnaire response:} 
-1 ||| More regular sleep, steadier mobility/location patterns, and increased phone engagement suggest less behavioral withdrawal and slightly improved daily functioning, which is more consistent with reduced feeling down/hopeless; the same signs of leaving home, visiting places, and engaging more by phone also suggest somewhat greater interest/pleasure rather than worsening anhedonia, though improvement appears partial not full.

%% file: sections/appx_features.tex
\section{Passive-sensing Features}
\label{app:features}
\begingroup
\footnotesize
\setlength{\tabcolsep}{3pt}
\renewcommand{\arraystretch}{1.1}
\begin{longtable}{
  >{\raggedright\arraybackslash}p{0.06\linewidth}
  >{\raggedright\arraybackslash}p{0.22\linewidth}
  >{\raggedright\arraybackslash}p{0.24\linewidth}
  >{\raggedright\arraybackslash}p{0.18\linewidth}
  >{\raggedright\arraybackslash}p{0.20\linewidth}}
\toprule
 & \multicolumn{2}{c}{\textbf{GLOBEM}} & \multicolumn{2}{c}{\textbf{College Experience Study}} \\
\cmidrule(lr){2-3}\cmidrule(lr){4-5}
\textbf{Domain} & \textbf{Raw Feature} & \textbf{Description} & \textbf{Raw Feature} & \textbf{Description} \\
\midrule
\endfirsthead
% \multicolumn{5}{l}{\textit{Table~\ref{tab:features} (continued)}}\\
\toprule
 & \multicolumn{2}{c}{\textbf{GLOBEM}} & \multicolumn{2}{c}{\textbf{College Experience Study}} \\
\cmidrule(lr){2-3}\cmidrule(lr){4-5}
\textbf{Domain} & \textbf{Raw Feature} & \textbf{Description} & \textbf{Raw Feature} & \textbf{Description} \\
\midrule
\endhead
\midrule
\multicolumn{5}{r}{\textit{Continued on next page}}\\
\endfoot
\bottomrule
\endlastfoot
% ===================== SLEEP =====================
Sleep
 & \feat{sumdurationasleepmain} & Total Sleep Time (min)
 & \feat{sleep_duration} & Sleep Duration (hr) \\
 & \feat{sumdurationinbedmain} & Time in Bed (min)
 & \feat{sleep_heathkit_dur} & Sleep Duration from HealthKit (hr) \\
 & \feat{firstbedtimemain} & First Bedtime
 & \feat{sleep_start} & Sleep Start Time \\
 & \feat{firstwaketimemain} & First Wake Time
 & \feat{sleep_end} & Sleep End Time \\
 & \feat{lastbedtimemain} & Last Bedtime & & \\
 & \feat{lastwaketimemain} & Last Wake Time & & \\
 & \feat{sumdurationawakemain} & Awake Time (min) & & \\
 & \feat{sumdurationtofallasleepmain} & Sleep Onset Latency (min) & & \\
 & \feat{sumdurationafterwakeupmain} & Time After Wake Up (min) & & \\
 & \feat{avgefficiencymain} & Sleep Efficiency (\%) & & \\
 & \feat{avgdurationasleepmain} & Average Sleep Duration (min) & & \\
 & \feat{avgdurationinbedmain} & Average Time in Bed (min) & & \\
 & \feat{avgdurationawakemain} & Average Awake Duration (min) & & \\
 & \feat{avgdurationtofallasleepmain} & Average Sleep Onset Latency (min) & & \\
 & \feat{avgdurationafterwakeupmain} & Average Time After Wake Up (min) & & \\
 & \feat{countepisodemain} & Number of Sleep Episodes & & \\
\midrule
% ===================== ACTIVITY =====================
Activity
 & \feat{avgsumsteps} & Average Steps
 & \feat{step_ep_0} & Number of Steps \\
 & \feat{mediansumsteps} & Median Steps
 & \feat{act_walking_ep_0} & Walking Duration (sec) \\
 & \feat{maxsumsteps} & Maximum Steps
 & \feat{act_running_ep_0} & Running Duration (sec) \\
 & \feat{minsumsteps} & Minimum Steps
 & \feat{act_on_bike_ep_0} & Time on Bike (sec) \\
 & & & \feat{act_in_vehicle_ep_0} & Time in Vehicle (sec) \\
 & & & \feat{act_still_ep_0} & Still Movement Duration (sec) \\
\midrule
% ===================== MOBILITY =====================
Mobility
 & \feat{totaldistance} & Total Distance Traveled (m)
 & \feat{loc_dist_ep_0} & Distance Traveled (m) \\
 & \feat{numberofsignificantplaces} & Number of Significant Places
 & \feat{loc_visit_num_ep_0} & Number of Locations Visited \\
 & \feat{timeathome} & Time at Home (min)
 & \feat{loc_home_dur} & Time at Home (hr) \\
 & \feat{duration_in_locmap_study} & Time at Study Locations (min)
 & \feat{loc_study_dur} & Time at Study Locations (hr) \\
 & \feat{duration_in_locmap_exercise} & Time at Exercise Locations (min)
 & \feat{loc_workout_dur} & Time at Workout Locations (hr) \\
 & \feat{duration_in_locmap_greens} & Time at Green Spaces (min)
 & \feat{loc_leisure_dur} & Time at Leisure Locations (hr) \\
 & \feat{radiusgyration} & Radius of Gyration (m)
 & \feat{loc_max_dis_from_campus_ep_0} & Maximum Distance from Campus (m) \\
 & \feat{numberlocationtransitions} & Number of Location Transitions
 & \feat{loc_food_dur} & Time at Food Locations (hr) \\
 & \feat{avgspeed} & Average Speed (km/h)
 & \feat{loc_health_dur} & Time at Health Locations (hr) \\
 & \feat{varspeed} & Speed Variance ((km/h)$^2$)
 & \feat{loc_other_dorm_dur} & Time at Others' Dorms (hr) \\
 & \feat{movingtostaticratio} & Ratio of Stationary Time to Total Location-Sensed Time
 & \feat{loc_self_dorm_dur} & Time at Own Dorm (hr) \\
 & \feat{outlierstimepercent} & Time at Non-Significant Places (\%)
 & \feat{loc_social_dur} & Time at Social Locations (hr) \\
 & \feat{timeattop1location} & Time at Most-Visited Place (min)
 & \feat{loc_worship_dur} & Time at Worship Locations (hr) \\
 & \feat{timeattop2location} & Time at 2nd Most-Visited Place (min) & & \\
 & \feat{timeattop3location} & Time at 3rd Most-Visited Place (min) & & \\
 & \feat{avglengthstayatclusters} & Average Stay Duration at Significant Places (min) & & \\
 & \feat{maxlengthstayatclusters} & Max.\ Stay Duration at Significant Places (min) & & \\
 & \feat{minlengthstayatclusters} & Min.\ Stay Duration at Significant Places (min) & & \\
 & \feat{stdlengthstayatclusters} & Std.\ of Stay Duration at Significant Places (min) & & \\
\midrule
% ===================== PHONE USAGE =====================
Phone Usage
 & \feat{countepisodeunlock} & Number of Unlock Episodes
 & \feat{unlock_num_ep_0} & Number of Phone Unlocks \\
 & \feat{sumdurationunlock} & Total Unlock Duration (min)
 & \feat{unlock_duration_ep_0} & Total Phone Unlock Duration (sec) \\
 & \feat{countepisodeunlock_locmap_home} & Number of Unlock Episodes at Home
 & \feat{loc_home_unlock_num} & Unlocks per Hour at Home \\
 & \feat{sumdurationunlock_locmap_home} & Total Unlock Duration at Home (min)
 & \feat{loc_home_unlock_duration} & Phone Use per Hour at Home (min) \\
 & \feat{avgdurationunlock} & Average Unlock Duration (min) & & \\
 & \feat{maxdurationunlock} & Max.\ Unlock Duration (min) & & \\
 & \feat{mindurationunlock} & Min.\ Unlock Duration (min) & & \\
 & \feat{stddurationunlock} & Std.\ of Unlock Duration (min) & & \\
 & \feat{firstuseafter00unlock} & Time of First Unlock After Midnight (min) & & \\
 & \feat{avgdurationunlock_locmap_home} & Average Unlock Duration at Home (min) & & \\
 & \feat{maxdurationunlock_locmap_home} & Max.\ Unlock Duration at Home (min) & & \\
 & \feat{mindurationunlock_locmap_home} & Min.\ Unlock Duration at Home (min) & & \\
 & \feat{stddurationunlock_locmap_home} & Std.\ of Unlock Duration at Home (min) & & \\
 & \feat{firstuseafter00unlock_locmap_home} & Time of First Unlock at Home After Midnight (min) & & \\
 & \feat{incoming_count} & Number of Incoming Calls & & \\
 & \feat{outgoing_count} & Number of Outgoing Calls & & \\
 & \feat{missed_count} & Number of Missed Calls & & \\
 & \feat{incoming_countmostfrequentcontact} & Incoming Calls from Most Frequent Contact & & \\
 & \feat{outgoing_countmostfrequentcontact} & Outgoing Calls from Most Frequent Contact & & \\
 & \feat{missed_countmostfrequentcontact} & Missed Calls from Most Frequent Contact & & \\
\end{longtable}
\endgroup